\documentclass[superscriptaddress,nofootinbib,a4paper,11pt]{article}
\usepackage{jheppub}
\usepackage{graphicx}
\usepackage{subcaption}
\usepackage{float}
\usepackage{amsmath}
\usepackage{amssymb}
\usepackage[utf8]{inputenc}
\usepackage{hyperref}
\usepackage{color}
\usepackage{slashed}
\allowdisplaybreaks

\newcommand{\be}{\begin{equation}}
\newcommand{\ee}{\end{equation}}

\begin{document}

\title{\boldmath Multi-scalar signature of self-interacting dark matter in the NMSSM and beyond}

\author[a]{Jinmian Li,}
\author[b,c]{Junle Pei,}
\author[a]{and Cong Zhang}

\affiliation[a]{College of Physics, Sichuan University, Chengdu 610065, China}
\affiliation[b]{CAS Key Laboratory of Theoretical Physics, Institute of Theoretical Physics, Chinese Academy of Sciences, Beijing 100190, China}
\affiliation[c]{School of Physical Sciences, University of Chinese Academy of Sciences,
No.~19A Yuquan Road, Beijing 100049, China}

\emailAdd{jmli@scu.edu.cn}
\emailAdd{peijunle@mail.itp.ac.cn}
\emailAdd{ZhangCong.phy@gmail.com}

\abstract{
This work studies the self-interacting dark matter (SIDM) scenario in the general NMSSM and beyond, where the dark matter is a Majorana fermion and the force mediator is a scalar boson.
An improved analytical expression for the dark matter (DM) self-interacting cross section which takes into account the Born level effects is proposed. 
Due to the large couplings and light mediator in SIDM scenario, the DM/mediator will go through multiple branchings if they are produced with high energy. 
Based on the Monte Carlo simulation of the showers in the DM sector, we obtain the multiplicities and the spectra 
of the DM/mediator from the Higgsino production and decay at the LHC for our benchmark points. 
}

\maketitle

\section{Introduction}

The weakly interacting massive particle being the dark matter (DM) candidate can successfully explain the large-scale structure of the universe from the galactic scale to the cosmological scale. 
However, there have been tensions between the N-body simulations of collisionless cold DM and astrophysical observations on small scale structure of the universe, dubbed small scale structure problem~\cite{Bullock:2017xww}. 
The issues can be resolved if the DM has strong self-interactions~\cite{Spergel:1999mh,Tulin:2017ara}.  
On the other hand, the self-interacting DM (SIDM) scenario is stringently constrained in high-velocity system such as galaxy clusters~\cite{Markevitch:2003at,Kahlhoefer:2013dca,Kaplinghat:2015aga}. 
A viable SIDM scenario requires the DM self-scattering cross section that is suppressed at higher velocity and increases toward smaller velocity. 
Such feature can be implemented if the DM self-interaction is mediated by a light ($\mathcal{O}(1)$ MeV) scalar or vector particle~\cite{Kaplinghat:2015aga,Ackerman:mha,Feng:2009mn,Buckley:2009in,Feng:2009hw,Loeb:2010gj,Aarssen:2012fx,Tulin:2013teo,Duerr:2018mbd,Duch:2019vjg,Kamada:2020buc,Kao:2020sqs}. 

The strongly interacting DM with scalar mediator can be naturally realized in the next-to-minimal supersymmetric standard model (NMSSM)~\cite{Franke:1995tc,Ellwanger:2009dp}. 
In the small $\lambda$ limit, the singlet sector (includes a CP-even scalar, a CP-odd pseudoscalar and a singlino) is nearly decoupled from the SM sector, such that they can be light while evade all experimental searches. 
The Peccei-Quinn limit of the NMSSM with light singlet sector has been studied in Ref.~\cite{Draper:2010ew}. However, the SIDM scenario requires large $\kappa$ coupling. In the $Z_3$ invariant NMSSM, the singlino mass is proportional to $\kappa / \lambda$, which is sizable in the limit of $\lambda \ll \kappa$. The general NMSSM is required to address the SIDM scenario~\cite{Wang:2014kja,Kang:2016xrm,Wang:2016lvj,Wang:2019jhb,Zhu:2021pad}. 
Meanwhile, the correct DM relic density can be set via thermal freeze-out with DM annihilating into scalar mediators. 

Given light mediator and DM, together with relatively large couplings in the SIDM scenario, the production of the mediator/DM at the electroweak scale or higher will be followed by copious emissions of the mediator in the collinear direction due to the logarithmic enhancement. This is analogous to the QED/QCD parton showering. The phenomenology of the DM showering in models with extra dark gauge group has been studied in Ref.~\cite{Cheung:2009su,Buschmann:2015awa,Park:2017rfb,Cohen:2020afv,Knapen:2021eip}, where the force mediators are massless dark photon/gluons (similar to the hidden valleys scenario~\cite{Strassler:2006im}). 
In the NMSSM, the mediator is the CP-even massive scalar and the DM is the massive Majorana fermion.
The properties (divergence behaviors) of the splitting function are quite different from the massless vector mediator ones. More detailed studies of the DM showering with massive states have been performed in Ref.~\cite{Zhang:2016sll,Chen:2018uii}, inspired by the studies for the electroweak showing~\cite{Chen:2016wkt,Bauer:2018xag}

In this paper, we focus on the SIDM scenario in the general NMSSM. 
An improved analytical estimation for the DM self-interacting cross section which takes into account the Born level effects is proposed. We will demonstrate that the analytical expression matches well with the numerical solution, in terms of calculations on the scanned points in the NMSSM. 
Two benchmark points (in the NMSSM) that address the small scale structure problem with correct relic density of SIDM and satisfy other phenomenological constraints will be provided for the first time. 
They are featured by light Higgsinos and large probability of $\phi \to  \phi \phi$ splitting. The multiplicity and the spectrum of the scalar mediator in the Higgsino production (and decay) process will be studied in detail, based on Monte Carlo simulation of the showers in the DM sector.
However, $\kappa$ can not be large in the NMSSM in order to implement correct DM relic density, which suppresses the DM splitting. Two benchmark points beyond the NMSSM with $\kappa=2.5$ and address the small scale structure issue are proposed. The DM splitting ($\chi \to \phi \chi$) becomes also significant in this case. 
A generic issue of SIDM scenario with simple DM sector is the tension between Big Bang Nucleosynthesis (BBN) limits and dark matter direct detection constraints~\cite{Kainulainen:2015sva,Blennow:2016gde,Kang:2016xrm,Kahlhoefer:2017umn,Barman:2018pez}. Possible solutions require extended particle content and couplings. And the collider signature of SIDM scenario will be highly depend on the forms of extensions. 

The rest of the paper is organized as follows. In Sec.~\ref{sec:simp}, we will discuss how to calculate the non-relativistic DM self-interaction cross section with analytic method and numerical method. In particular, an improved analytical estimation for the DM self-interacting cross section is proposed. In Sec.~\ref{sec:nmssmsimp}, focusing on the general NMSSM model, we study the parameter space that addresses the small scale structure problem and is consistent with other phenomenological constraints. 
The showering of the DM and the mediator is discussed in Sec.~\ref{sec:splitting}. 
Finally, in Sec.~\ref{sec:simplhc}, we discuss the possible LHC signatures of the SIDM in the NMSSM and beyond, based on four selected benchmark points.

\section{Self-interacting dark matter - scalar mediator} \label{sec:simp}

The scattering between DMs ($\chi$) with a scalar mediator ($\phi$) in the non-relativistic limit is controlled by an attractive Yukawa potential
\begin{equation}\label{Vr}
V(r)=-\alpha_\chi\frac{e^{-m_\phi r}}{r}~,
\end{equation}
where $\alpha_\chi=\kappa^2 / (2\pi)$ and $\sqrt{2}\kappa$ is the coupling of $\phi \chi \chi$. The scattering amplitude is
\begin{equation}\label{ftheta}
f(\theta)=\frac{1}{k}\sum_{l=0}^{\infty}(2l+1)e^{i\delta_l}P_l(\cos\theta)\sin\delta_l~,
\end{equation}	
where $\delta_l$ is the phase shift for a partial wave $l$. It can be obtained by solving the Schr{\" o}dinger equation for given potential $V(r)$,  and $k=m_{\chi} v / 2$ with $v$ being the relative velocity between the DMs in the scattering. 

To describe the scattering between distinguishable particles, the transfer cross section $\sigma_\mathrm{T}$ and the viscosity cross section $\sigma_\mathrm{V}$ are usually used~\cite{PhysRevA.60.2118}, which are defined as
\begin{align}
\sigma_{\mathrm{T}}=\int d \Omega(1-\cos \theta) \frac{d \sigma}{d \Omega}~, \quad \sigma_{\mathrm{V}}=\int d \Omega \sin ^{2} \theta \frac{d \sigma}{d \Omega}~.
\end{align}
The relation between the differential scattering cross section $d \sigma / d \Omega$ and the phase shift $\delta_l$ is given by
\begin{align}
\frac{d \sigma}{d \Omega}=\frac{1}{k^{2}}\left|\sum_{\ell=0}^{\infty}(2 \ell+1) e^{i \delta_{\ell}} P_{\ell}(\cos \theta) \sin \delta_{\ell}\right|^{2}.
\end{align}
So it can be calculated that 
\begin{align}
\sigma_{\mathrm{T}}  &=\frac{4 \pi}{k^{2}} \sum_{\ell=0}^{\infty}(\ell+1) \sin ^{2}\left(\delta_{\ell+1}-\delta_{\ell}\right)~, \\
\sigma_{\mathrm{V}}  &=\frac{4 \pi}{k^{2}} \sum_{\ell=0}^{\infty} \frac{(\ell+1)(\ell+2)}{2 \ell+3} \sin ^{2}\left(\delta_{\ell+2}-\delta_{\ell}\right)~.
\end{align}

When describing the scattering between identical particles, only the viscosity cross section $\sigma_\mathrm{V}$ is useful~\cite{Colquhoun:2020adl}. For Majorana fermion DM scattering, the spatial wave function should be symmetric (antisymmetric) when the total spin of the DM pair is 0 (1). Thus, $\sigma_\mathrm{V}$ is replaced by $\sigma_\mathrm{VS}$ and $\sigma_\mathrm{VA}$ in the symmetric case and antisymmetric case, respectively, which are
\begin{align}
\sigma_\mathrm{VS}&=\frac{1}{2}\int d\Omega \sin^2\theta|f(\theta)+f(\pi-\theta)|^2=\frac{8\pi}{k^2}\sum_{l=0}^{\infty}\frac{(2l+1)(2l+2)}{(4l+3)}\sin^2(\delta_{2l+2}-\delta_{2l})~,\label{svs}\\
\sigma_\mathrm{VA}&=\frac{1}{2}\int d\Omega \sin^2\theta|f(\theta)-f(\pi-\theta)|^2 =\frac{8\pi}{k^2}\sum_{l=0}^{\infty}\frac{(2l+2)(2l+3)}{(4l+5)}\sin^2(\delta_{2l+3}-\delta_{2l+1})~. \label{sva}
\end{align}
Note that a symmetry factor $1/2$ is inserted in integrals to avoid the double-counting in scattering of two identical particles. In the following analysis, we assume the DMs participating the scattering are unpolarized and refer to $\sigma_\mathrm{V}$ as the one averaging over the all spins:    
\begin{equation}\label{sv}
\sigma_\mathrm{V}=\frac{1}{4}\sigma_\mathrm{VS}+\frac{3}{4}\sigma_\mathrm{VA}~.
\end{equation}

To solve the Schr{\" o}dinger equation, it is useful to define the new variables:
\begin{equation}\label{ab}a\equiv \frac{v}{2\alpha_\chi}~,~~~~~~~~b\equiv \frac{\alpha_\chi m_\chi}{m_\phi}~.
\end{equation}
In general, there is no analytic solution to the Schr{\" o}dinger equation with the Yukawa potential on the $a-b$ plane. However, in the Born regime where $b \lesssim 1$, computed perturbatively in $\alpha_\chi$, the scattering  amplitude can be expressed as
\begin{equation}\label{scamb}
f(\theta)\approx
\frac{2\alpha_\chi m_\chi}{m^2_\phi+4k^2\sin^2\frac{\theta}{2}}
\end{equation}
at the leading order, from which we can calculate the leading order DM scattering cross sections as
\begin{align}
\sigma^{\text{Born}}_\mathrm{T}&=\frac{\pi}{2a^2 k^2}\left(\ln\left( 1+4t^2\right)  -\frac{4t^2}{1+4t^2}\right)~,\label{st}\\
\sigma^{\text{Born}}_\mathrm{VS}&=\frac{\pi}{a^2 k^2}\left(\frac{(1+4t^2+8t^4)\ln(1+4t^2)}{4t^2+8t^4}-1 \right)~,\label{svsb}\\
\sigma^{\text{Born}}_\mathrm{VA}&=\frac{\pi}{a^2 k^2}\left(\frac{(3+12t^2+8t^4)\ln(1+4t^2)}{4t^2+8t^4}-3 \right)~,\label{svab}
\end{align}
where $t=ab$.
Besides, in the Born regime, we can also use~\cite{taylor2006} 
\begin{equation}\label{phsb}
e^{i\delta_l}\sin\delta_l=\frac{1}{a}\int_{0}^{\infty}dx x e^{-\frac{x}{t}}j_l^2(x)
\end{equation}
to estimate the $\delta_l$ at the leading order, where $j_l$ is the spherical Bessel function. 

In the quantum regime where $t \lesssim 1$,  the s-wave scattering is dominant, {\it i.e.} $|\delta_{0}|\gg |\delta_{l}|$ for $l>0$.
By taking the Hulth{\' e}n approximation, $\delta_{0}$ for the attractive Yukawa potential is given by~\cite{Tulin:2013teo}
\begin{align}
\delta_{0}^{\text{Hulth{\' e}n}}= \arg\left(\frac{i\Gamma(\lambda_+ + \lambda_- -2)}{\Gamma(\lambda_+)\Gamma(\lambda_-)} \right)
\end{align}
with
\begin{align}
\lambda_\pm=1+iac\pm\sqrt{c-a^2c^2}~,
\end{align}
where $c\approx {b}/{1.6}$. So we can use~\cite{Tulin:2013teo,Colquhoun:2020adl}
\begin{align}
\sigma_\mathrm{T}^{\text{Quan}}&= \frac{4\pi}{k^2}\sin^2\left( \delta_{0}^{\text{Hulth{\' e}n}}\right)~, \\
\sigma_\mathrm{VS}^{\text{Quan}}&= \frac{8\pi}{k^2}\sin^2\left( \delta_{0}^{\text{Hulth{\' e}n}}\right)~, \\
\sigma_\mathrm{VA}^{\text{Quan}}&= 0~, 
\end{align} 
as approximate expressions in the quantum regime.

Recent study in Ref.~\cite{Colquhoun:2020adl} provides the analytic approximations of $\sigma_\mathrm{T}$, $\sigma_\mathrm{VS}$, and $\sigma_\mathrm{VA}$ for both attractive and repulsive Yukawa potentials in the semi-classical regime where $t \gtrsim 1$. The cross sections in this regime are strongly depend on $\beta=1/ (2a^2b)$. For the attractive Yukawa potential, the expressions for cross sections are summarized as
\begin{align}
\sigma_{\mathrm{T}}^{\text {Clas}}&=\frac{\pi}{m_{\phi}^{2}} \times\left\{\begin{array}{ll}
2 \beta^{2} \zeta_{\frac{1}{2}}(t, \beta) & \beta \leq 0.2 \\
2 \beta^{2} \zeta_{\frac{1}{2}}(t, \beta) e^{0.64(\beta-0.2)} & 0.2<\beta \leq 1 \\
4.7 \log (\beta+0.82) & 1<\beta<50 \\
2 \log \beta(\log \log \beta+1) & \beta \geq 50
\end{array}\right.~,\\
\sigma_{\mathrm{VS/A}}^{\text{Clas}}&=\frac{\pi}{m_{\phi}^{2}} \times\left\{\begin{array}{ll}
4\beta^{2}   \zeta_{n}(t, 2 \beta)  & \beta \leq 0.1 \\
4\beta^{2}  \zeta_{n}(t, 2 \beta) e^{0.67(\beta-0.1)} & 0.1<\beta \leq 0.5 \\
2.5 \log (\beta+1.05) & 0.5<\beta<25 \\
\frac{1}{2}\left(1+\log \beta-\frac{1}{2 \log \beta}\right)^{2} & \beta \geq 25
\end{array}\right.~,
\end{align}
with
\begin{align}
\zeta_{n}(t, \beta) &=\frac{\max (n, \beta t)^{2}-n^{2}}{2 t^{2} \beta^{2}}+\eta\left(\frac{\max (n, \beta t)}{t}\right)~, \\
\eta(x) &=x^{2}\left[-K_{1}(x)^{2}+K_{0}(x) K_{2}(x)\right] ~,
\end{align}
where $n=\frac{1}{2}\left(n=\frac{3}{2}\right)$ for $\sigma_\mathrm{VS}^{\text{Clas}}\left(\sigma_\mathrm{VA}^{\text{Clas}}\right)$  and $K_i~(i=0,1,2)$ stands for the modified Bessel functions of the second kind.

\begin{figure}[htb]
	\begin{center}
	\begin{minipage}{0.9\textwidth}
		\includegraphics[width=0.32\textwidth]{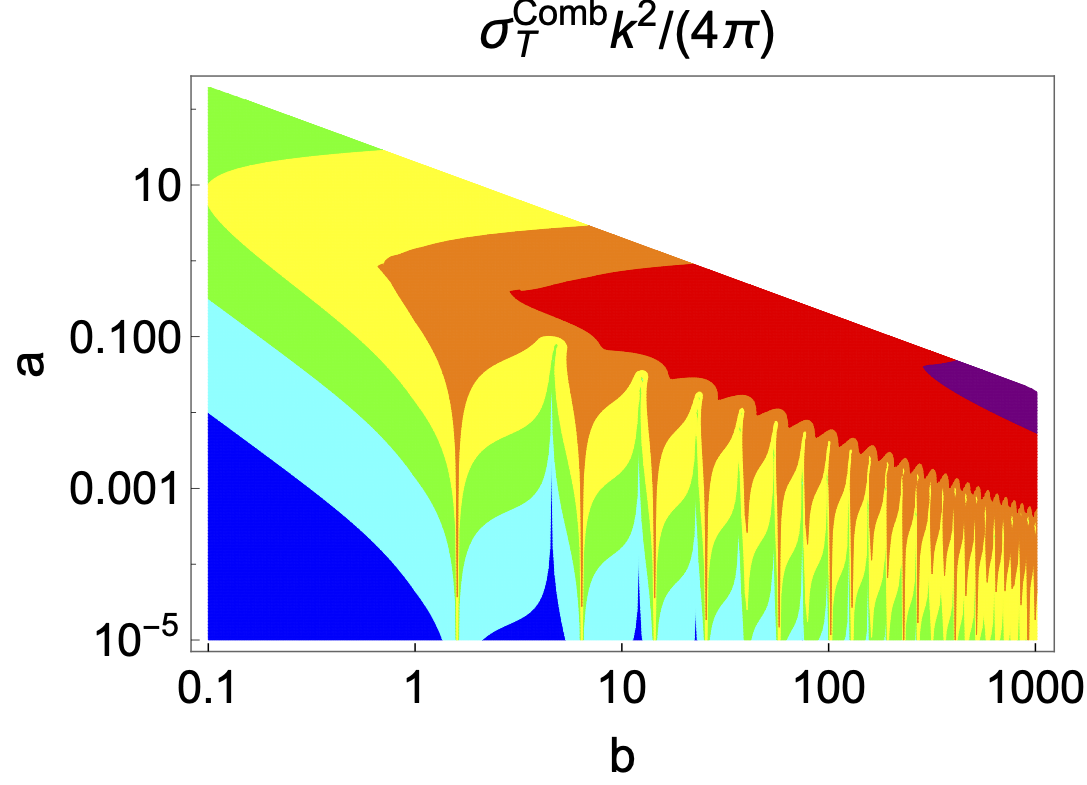}
		\includegraphics[width=0.32\textwidth]{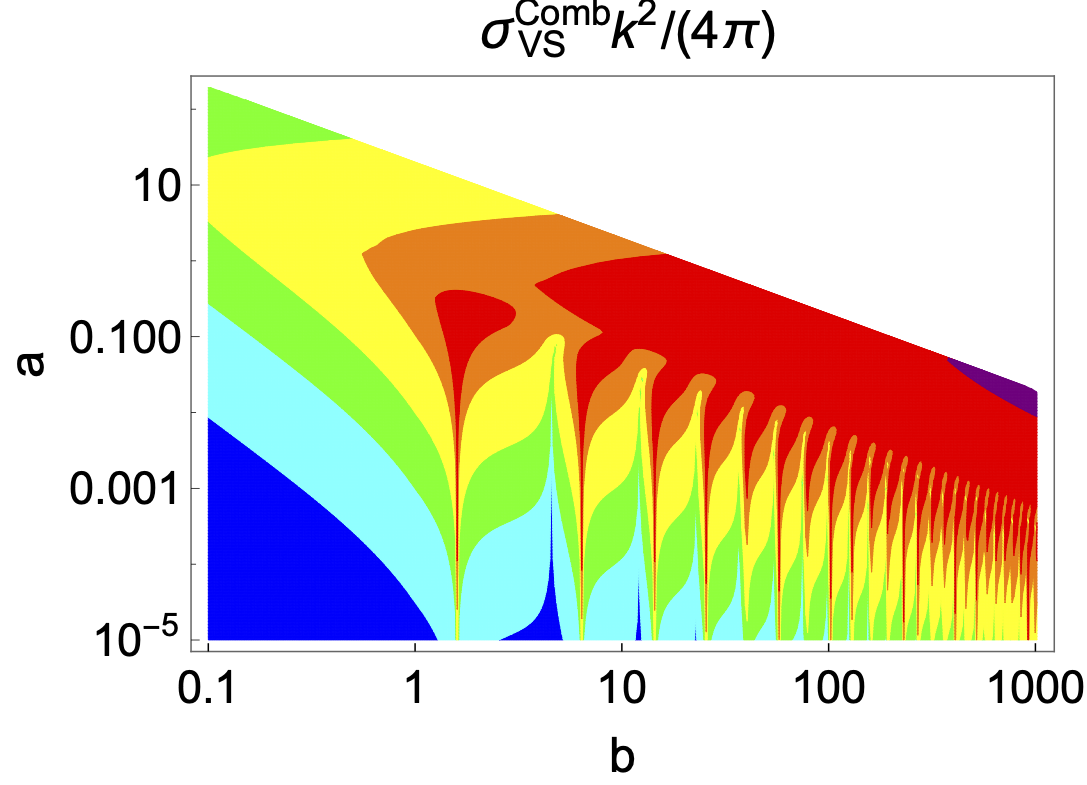}
		\includegraphics[width=0.32\textwidth]{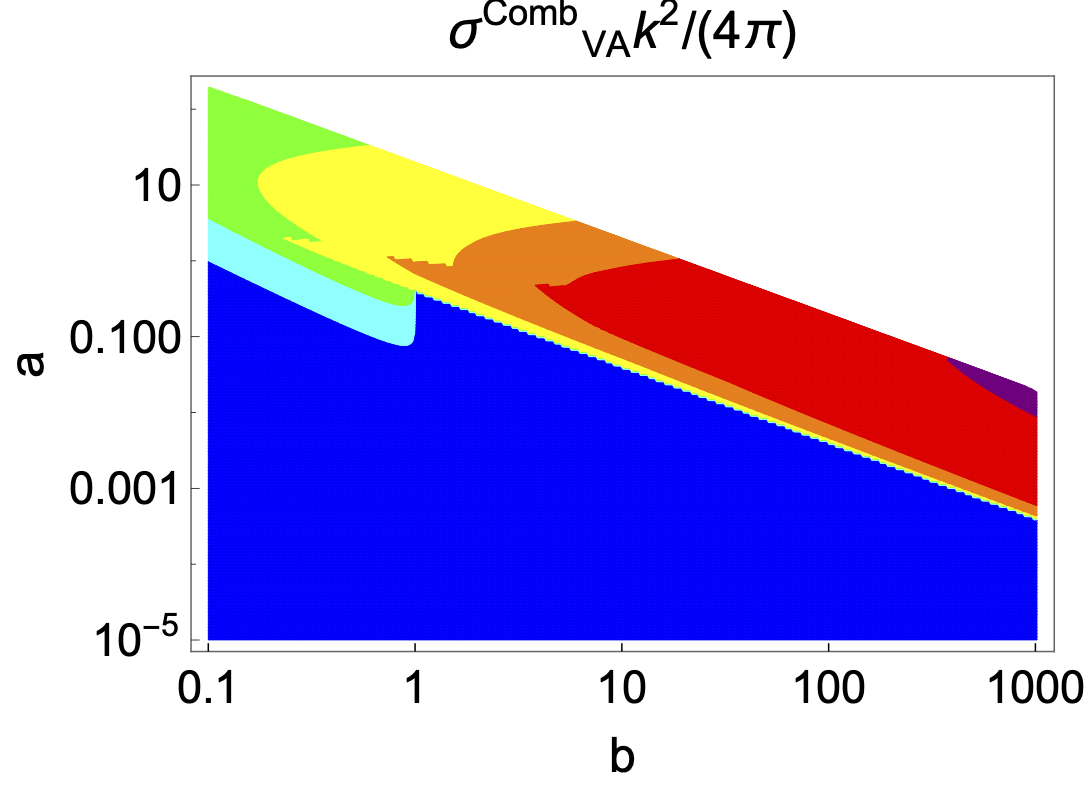}\\   
	        \includegraphics[width=0.32\textwidth]{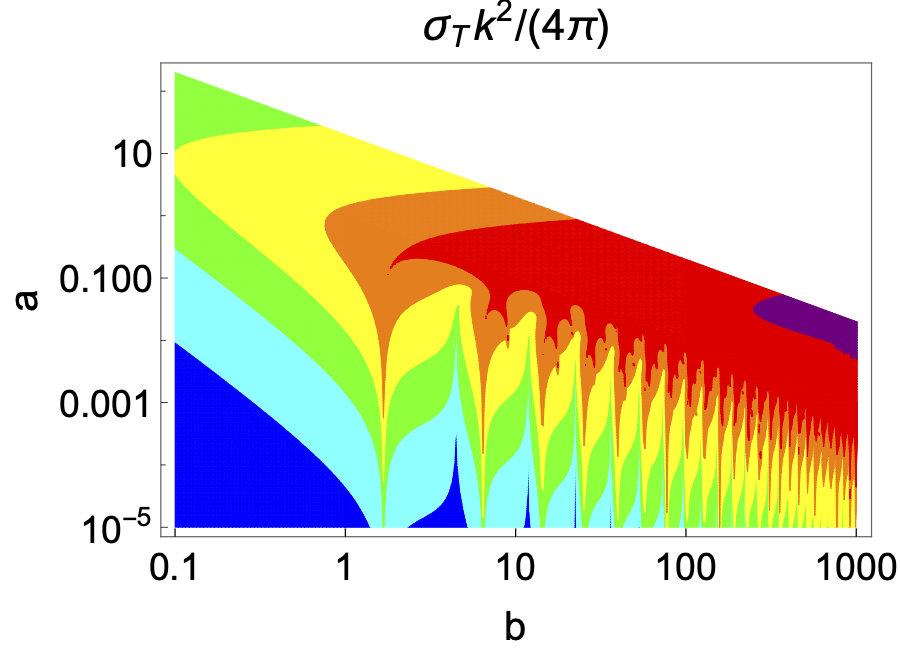}
		\includegraphics[width=0.32\textwidth]{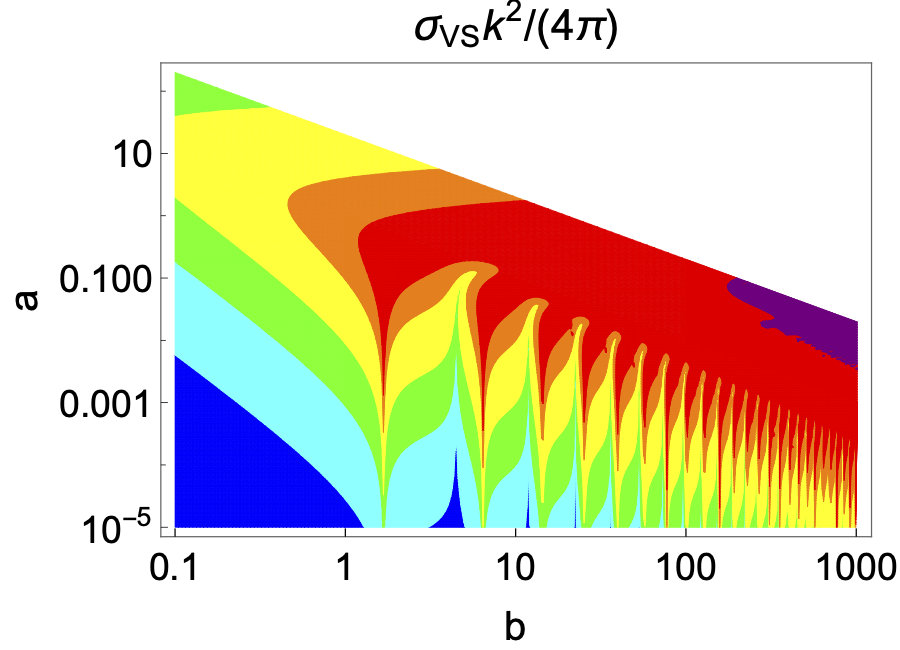}
		\includegraphics[width=0.32\textwidth]{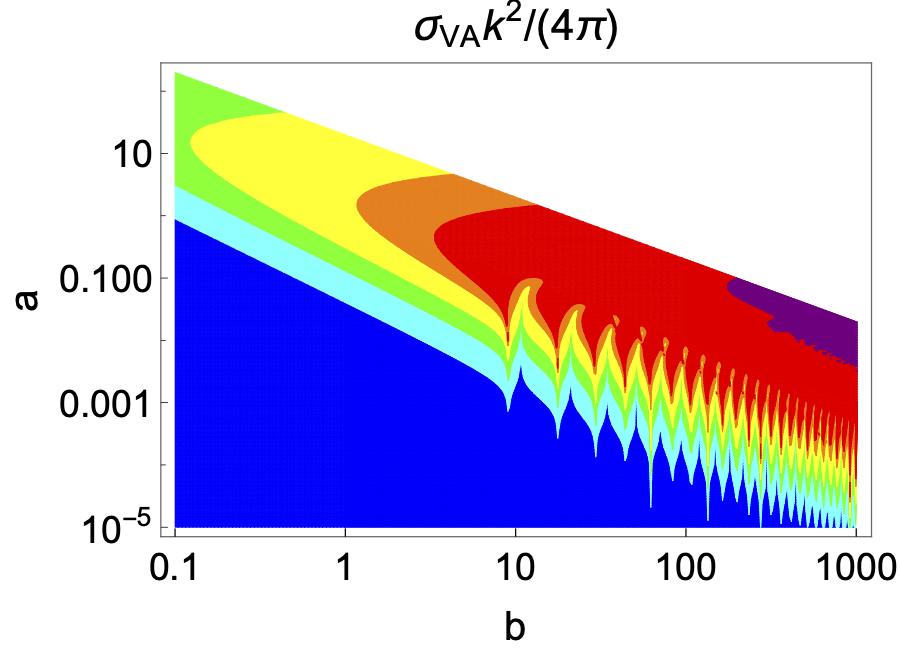}
	\end{minipage}%
	\begin{minipage}{0.1\textwidth}
	\includegraphics[width=0.95\textwidth]{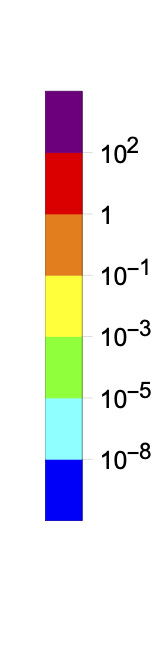}
	\end{minipage}
	\end{center}
	\caption{\label{fig:sigma_ana} The transfer cross section $\sigma_\mathrm{T}$ ($\times k^2/(4\pi)$), the symmetric viscosity cross section $\sigma_{\mathrm{VS}}$ ($\times k^2/(4\pi)$) and the asymmetric viscosity cross section $\sigma_{\mathrm{VA}}$ ($\times k^2/(4\pi)$) on the $a$-$b$ plane, calculated by Eq.~\ref{eq:sigma_comb} (upper panels) and calculated by numerically solving the Schr{\" o}dinger equation (low panels). }
\end{figure}

The $\sigma_{\mathrm{T}}$ and the spin averaged $\sigma_{\mathrm{V}}$ on the whole $a-b$ plane can be obtained by combining the expressions above. They are

\begin{align}
\sigma_{\mathrm{i}}^{\text{Comb}}=\left\{\begin{array}{ll}
\sigma_{\mathrm{i}}^{\text{Born}}  & b \leq 0.1 \\
\frac{1-b}{0.9}\sigma_{\mathrm{i}}^{\text{Born}} +\frac{b-0.1}{0.9}\sigma_{\mathrm{i}}^{\text{Q-C}} & 0.1<b < 1 \\
\sigma_{\mathrm{i}}^{\text{Q-C}} & b \geq 1
\end{array}\right.,~~~~~~~~i=\text{T},\text{VS},\text{VA},\text{V}, \label{eq:sigma_comb}
\end{align}  
where
\begin{align}
&\sigma_{\mathrm{i}}^{\text{Q-C}}=\left\{\begin{array}{ll}
\sigma_{\mathrm{i}}^{\text{Quan}}  & t \leq 0.4 \\
\frac{1-t}{0.6}\sigma_{\mathrm{i}}^{\text{Quan}} +\frac{t-0.4}{0.6}\sigma_{\mathrm{i}}^{\text{Clas}} & 0.4<t < 1 \\
\sigma_{\mathrm{i}}^{\text{Clas}} & t \geq 1
\end{array}\right.,~~~~~~~~i=\text{T},\text{VS},\text{VA},\text{V},\\
&\sigma_{\mathrm{V}}^{j}=\frac{1}{4}\sigma_{\mathrm{VS}}^{j}+\frac{3}{4}\sigma_{\mathrm{VA}}^{j}~,~~~~~~~~j=\text{Born},\text{Quan},\text{Clas}.
\end{align}  
It should be noted that our final combined expressions $\sigma_{\mathrm{i}}^{\text{Comb}}~(i=\text{T},\text{V})$ are different from the analytic expressions proposed in Ref.~\cite{Colquhoun:2020adl}. The effects of the $\sigma_{\mathrm{i}}^{\text{Born}}$ have been taken into account in our case. The shapes of the transfer cross section $\sigma^{\text{Comb}}_\mathrm{T}$ and the viscosity cross sections $\sigma^{\text{Comb}}_{\mathrm{VS}}$ and $\sigma^{\text{Comb}}_{\mathrm{VA}}$ (all are scaled by a factor of $\frac{k^2}{4\pi}$) are illustrated in upper panels of Fig.~\ref{fig:sigma_ana}. 

Meanwhile, following the method proposed in Ref.~\cite{Tulin:2013teo}, the Schr{\" o}dinger equation with Yukawa potential can be solved numerically. In the lower panels of Fig.~\ref{fig:sigma_ana}, we show the $\sigma_\mathrm{\text{T}}k^2/(4\pi)$, $\sigma_\mathrm{\text{\text{VS}}}k^2/(4\pi)$, and $\sigma_\mathrm{\text{VA}}k^2/(4\pi)$ obtained with the numerical calculation. Comparing the analytical and numerical results, we can observe that the analytical estimations match reasonably well with those from numerical calculations. A more specific comparison based on the scanned parameter points in the NMSSM will be given later in this work. 
As for the asymmetric viscosity cross section, the analytical calculation in the quantum regime raises zero, which is in consistent with the numerical calculation as most points in the quantum regime give $\sigma_{\mathrm{VA}}k^2/(4\pi) \lesssim 10^{-8}$ and it is much smaller than $\sigma_\mathrm{\text{T}}k^2/(4\pi)$ and $\sigma_\mathrm{\text{\text{VS}}}k^2/(4\pi)$.

\section{The NMSSM and SIDM} \label{sec:nmssmsimp}

In this section, we will realize the SIDM scenario in the NMSSM, considering the phenomenological constraints. In particular, the relic density and DM direct detection constraints will be discussed in detail. 

\subsection{The NMSSM - masses and couplings of the singlet sector}  \label{sec:nmssm}
The NMSSM is a well-motivated extension of the MSSM by a gauge singlet chiral superfield $\hat{S}$. 
Its most general superpotential is:
\begin{align}
W = W_{\text{Yukawa}} + (\mu + \lambda \hat{S}) \hat{H}_u \hat{H}_d + \xi_F \hat{S} +\frac{1}{2} \mu^\prime \hat{S}^2 + \frac{\kappa}{3} \hat{S}^3~,~
\end{align}
where $W_{\text{Yukawa}}$ describes the Yukawa couplings of quark and lepton superfields. We choose $\mu=0$ in this work, following the convention in Ref.~\cite{Ellwanger:2009dp} and NMSSMtools~\cite{Ellwanger:2004xm,Ellwanger:2005dv}.
The corresponding soft SUSY breaking terms are 
\begin{align}
\mathcal{L}_{\text{soft}} =  \mathcal{L}_{\text{soft}}^{\text{MSSM}} - m^2_S |S|^2 - (\lambda A_\lambda H_u H_d S + \frac{1}{3} \kappa A_\kappa S^3 + m^2_H H_u H_d + \frac{1}{2} m_S^{\prime 2} S^2 + \xi_S S +{\it h.c.} ) ~.~
\end{align}

After the electroweak symmetry breaking, the scalar fields $H_u$, $H_d$ and $S$ obtain vacuum expectation values $v_u$, $v_d$ and $s$, respectively. 
The elements of the CP-even scalar mass matrix square $\mathcal{M}^2_S$ in the basis $(H_{d}, H_{u},S)$ can be written as follows (only those relevant to the singlet are shown): 
\begin{align}
\mathcal{M}^2_{S,13} &= \lambda (2 \mu_{\text{eff}} v_d - v_u(B_{\text{eff}} +\kappa s + \mu^\prime  ) ) ~,~ \nonumber \\
\mathcal{M}^2_{S,23} &= \lambda (2 \mu_{\text{eff}} v_u - v_d(B_{\text{eff}} +\kappa s + \mu^\prime  ) ) ~,~ \nonumber  \\
\mathcal{M}^2_{S,33} &= \lambda \frac{v_u v_d}{s} (A_\lambda + \mu^\prime) +\kappa s (A_\kappa + 4 \kappa s + 3 \mu^\prime) - (\xi_S +\xi_F \mu^{\prime})/s ~,~
\end{align} 
with $\mu_{\text{eff}}=\lambda s$ and $B_{\text{eff}}=A_\lambda +\kappa s$. 
We denote the mass eigenstates of the mixing of the CP-even scalars from $H_u$, $H_d$, and $S$ as $H_i$ ($i=1,2,3$) satisfying $m_{H_1}<m_{H_2}<m_{H_3}$, so $H_1$ is the scalar mediator $\phi$.
And the mass matrix of neutralino in the basis $(\tilde{H}_d, \tilde{H}_u, \tilde{S})$ is read as
\begin{equation}
M_{\chi}= 
\begin{pmatrix}
0 & -\mu_{\text{eff}} & -\lambda v_u \\ 
-\mu_{\text{eff}} & 0  & -\lambda v_d\\
-\lambda v_u & -\lambda v_d & 2\kappa s + \mu^{\prime}
\end{pmatrix} ~.~
\end{equation}
The mass eigenstates of the neutralinos are denoted as $\chi_i$ ($i=1,2,3$) satisfying $m_{\chi_1}<m_{\chi_2}<m_{\chi_3}$, so $\chi_1$ is the DM $\chi$.

In the limit where the mixing between the singlet scalar ($S=\frac{s+S_h+iS_a}{\sqrt{2}}$) and doublet Higgs fields ($H_u$, $H_d$), as well as the mixing between singlino ($\tilde{S}$) and Higgsinos ($\tilde{H}_u$, $\tilde{H}_d$) are small, we have $\phi=S_h$ and $\chi=\tilde{S}$ with masses given by 
\begin{align}
&m_{S_h}  = \sqrt{\mathcal{M}^2_{S,33}}~, \\
&m_{\tilde{S}}  =  2\kappa s + \mu^{\prime}~.
\end{align}	
And the couplings in the singlet sector can be evaluated as~\cite{Ellwanger:2009dp}: 
\begin{align} 
&V_{S S S} = -i4\kappa\left(  6\kappa s + A_\kappa +3\mu^\prime\right)~,  \\ 
&V_{S \tilde{S} \tilde{S}} =  -i2\kappa~. 
\end{align}

\subsection{SIDM in the NMSSM}
There are three parameters that control the DM self-scattering, $m_\chi$, $m_\phi$ and the coupling $V_{\phi \chi \chi}$. In the NMSSM with nearly decoupled singlet sector, we have $m_\chi=m_{\tilde{S}}$, $m_\phi = m_{S_h}$ and $V_{\phi \chi \chi}=V_{S \tilde{S} \tilde{S}}/\sqrt{2} = -i\sqrt{2} \kappa$. 
Since the singlino-like DM is a Majorana fermion, the spin averaged viscosity cross section $\sigma_\mathrm{V}$ is used to describe the non-relativistic scattering between self-interacting DMs and account for the small scale structure of the universe as have been discussed in Sec.~\ref{sec:simp}. 
The pseudoscalar mediator can not induce large DM self-interaction~\cite{Kahlhoefer:2017umn}, as there will be further contributions to the Yukawa potential scaling as $m^2_\phi/ m^2_\chi \times e^{-m_\phi r}/r^n$ with $n \ge 2$.

\begin{figure}[thb]
		\begin{center}
		\begin{minipage}{0.8\textwidth}
			\includegraphics[width=0.49\textwidth]{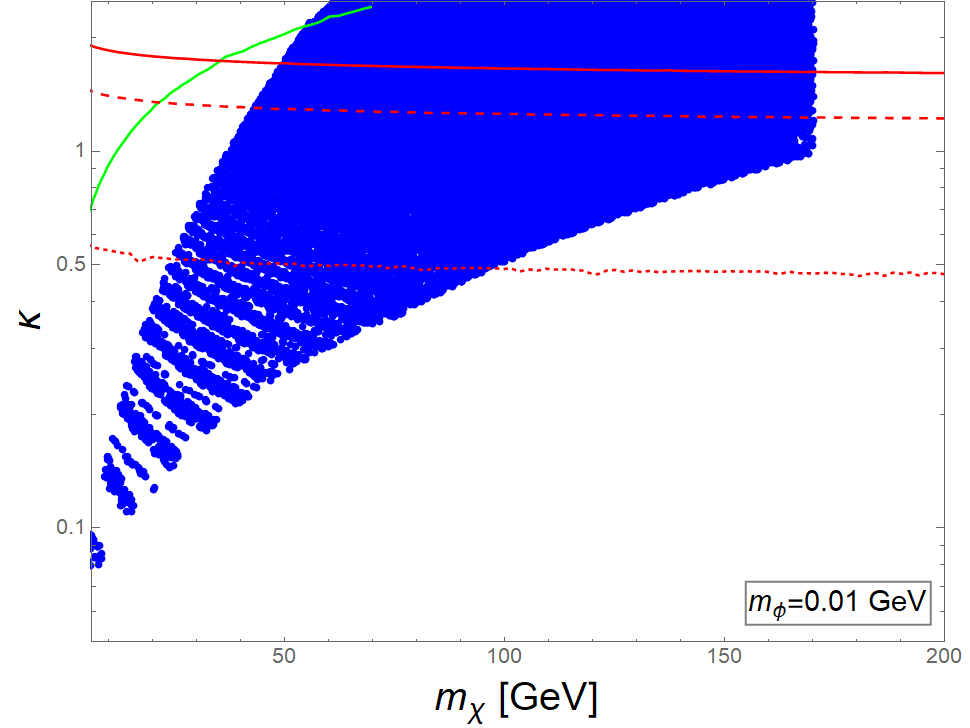}
			\includegraphics[width=0.49\textwidth]{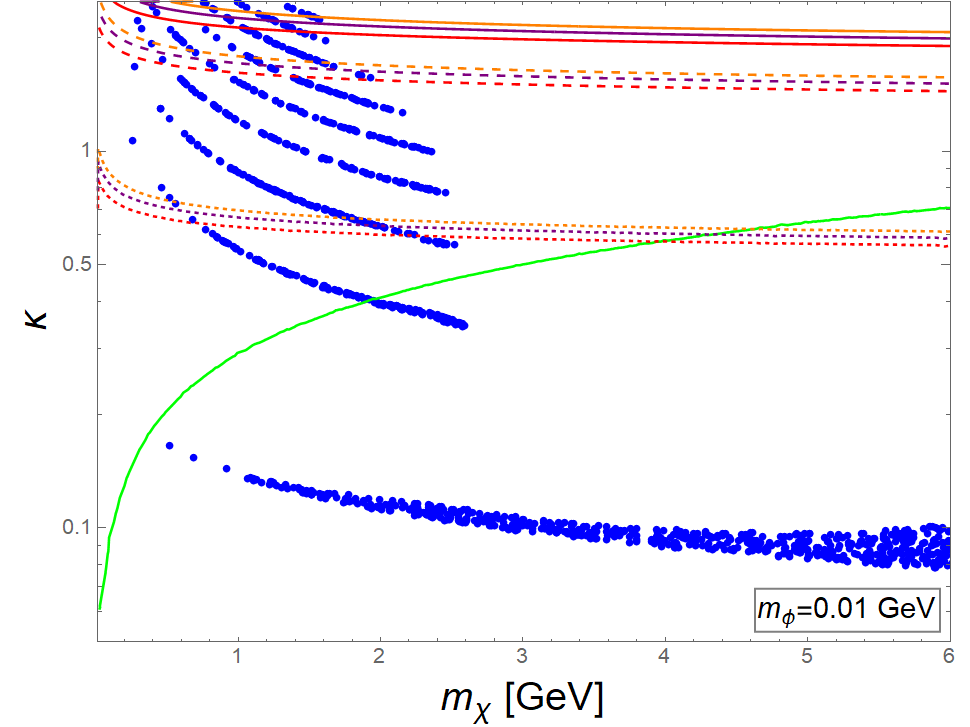}\\   
			\includegraphics[width=0.49\textwidth]{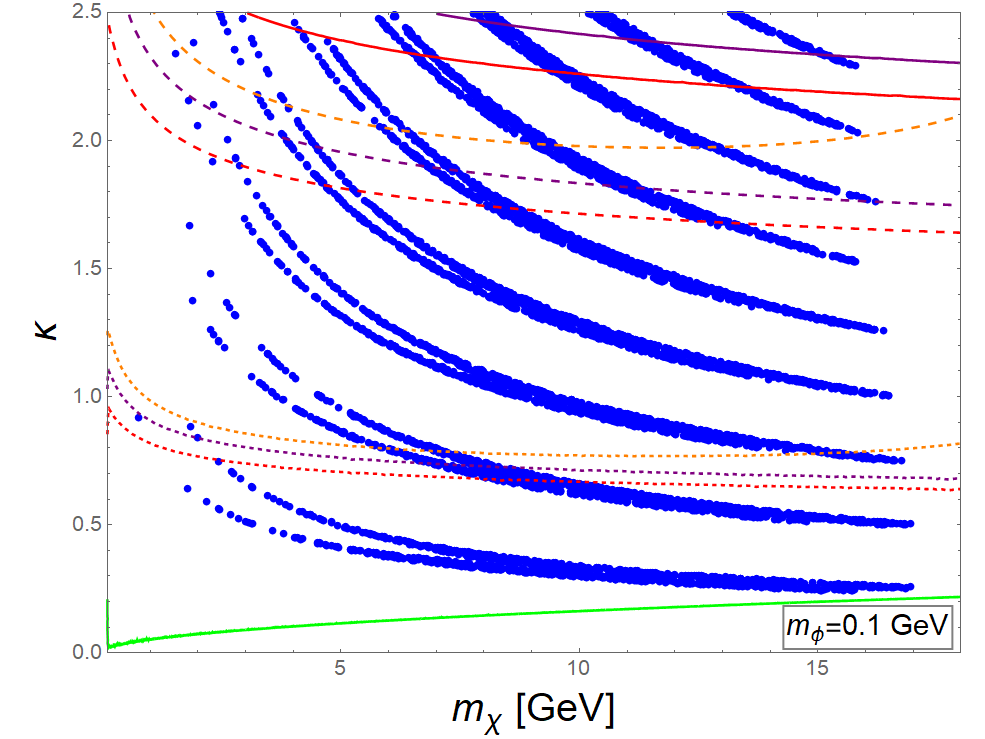}
			\includegraphics[width=0.49\textwidth]{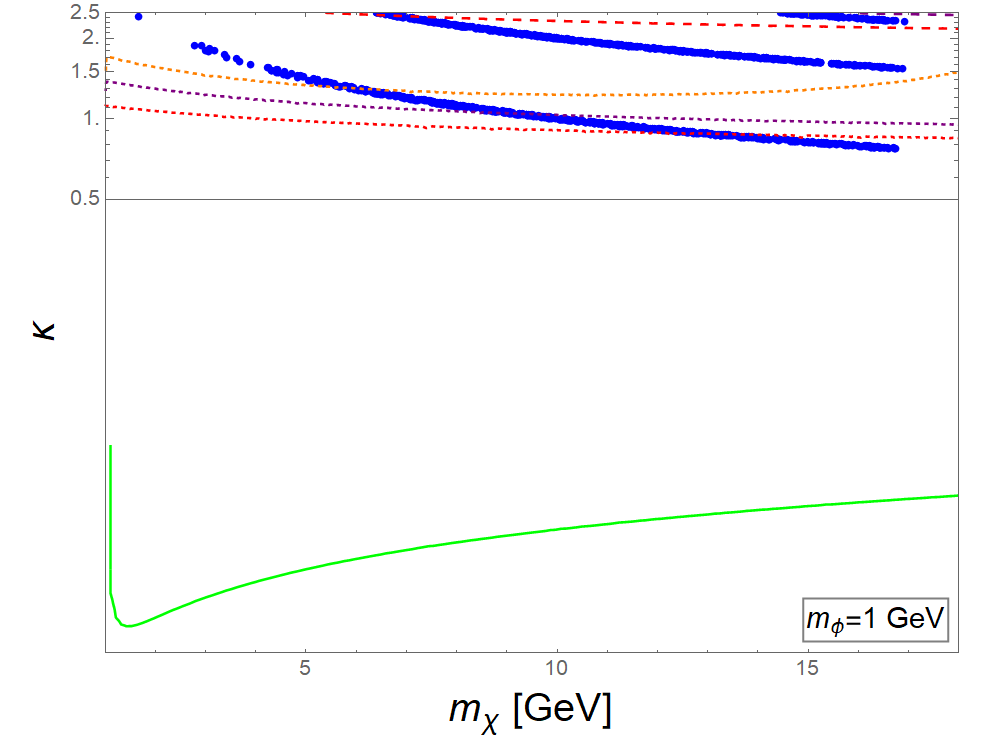}
		\end{minipage}%
		\begin{minipage}{0.2\textwidth}
			\includegraphics[width=0.99\textwidth]{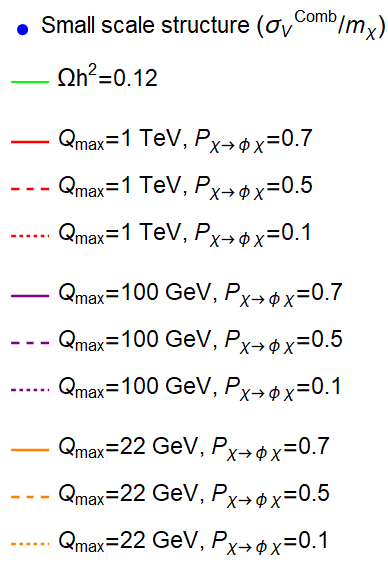}
		\end{minipage}
	\end{center}
	\caption{\label{fig:sigmaom}  Dots represent the region, where the small scale structure problem can be addressed, according to analytic formulas  $\sigma_{\text{V}}^{\text{Comb}}/m_\chi$ (blue). The green lines correspond to the parameters that give the correct relic density. The red, purple and orange contours indicate the splitting probabilities of $\chi \to \phi \chi$ for initial $\chi$ energies ($Q_{\max}$) 1 TeV, 100 GeV and 22 GeV, respectively. More details for the splitting will be discussed in Sec.~\ref{sec:splitting}. Three different masses of the scalar mediator are considered.}
\end{figure}

According to the simulations in Ref.~\cite{Tulin:2013teo,Laha:2013gva}, $\sigma_\mathrm{V}/m_\chi\sim 1-10~\text{cm}^2/\text{g}$ on dwarf scales (the characteristic velocity is 10 km/s) is needed to solve the core-vs-cusp and too-big-to-fail problems, while the constraints on 
cluster scales (the characteristic velocity is 1000 km/s) require $\sigma_\mathrm{V}/m_\chi\sim 0.1-1~\text{cm}^2/\text{g}$. In Fig.~\ref{fig:sigmaom}, we show the regions on the $\kappa$-$m_\chi$ plane, where the self-scattering cross section of SIDM is consistent with the simulations on dwarf and cluster scales. The mass of the scalar mediator $m_{\phi}$ is chosen to be 0.01 GeV,  0.1 GeV, and 1 GeV, respectively. 
In the upper left panel where $m_\phi=0.01$ GeV, the blue points which can address the small scale structure problems are in the classical regime on cluster scales ($v\sim 1000$ km/s) as they satisfy $t>1$ when $m_\chi>6$ GeV. And they are in the quantum regime on dwarf scales ($v\sim 10$ km/s) because $t<1$ is fulfilled when $m_\chi<600$ GeV. 
Further calculations show that the blue region shrinks toward smaller $m_\chi$ when we increase $m_\phi$ due to the constraints on cluster scale and disappears when $m_\phi$ is large than 0.07 GeV.
As heavy DM scenarios are stringently constrained by DM direct detection experiments, we will focus on DM mass $\lesssim 20$ GeV in parameter scan. 
In the upper right panel where $m_\phi=0.01$ GeV, the blue points are in the quantum regime on both dwarf and cluster scales.
In the lower left panel where $m_\phi=0.1$ GeV, the blue points come from the quantum regime (on  dwarf and cluster scales). The multi-band structure of the blue region corresponds to the quantum regime in the $a-b$ plane in Fig.~\ref{fig:sigma_ana}, which contains many  peaks and valleys. It is shown that there is an upper limit on $m_\chi$, which can be understood by that $t=v m_\chi /(2m_\phi)<1$ ($m_\chi<2m_\phi / v$) is satisfied in the quantum regime.
In the lower right panel where $m_\phi=1$ GeV,  the blue points also come from the quantum regime (on dwarf and cluster scales) and thus also have an upper limit on $m_\chi$. 
The blue belts in the $m_\phi=1$ GeV case are narrower than those in the $m_\phi=0.1$ GeV case because larger value of $\sigma_{\text{V}}$ is required to explain the small scale structure for heavier $m_\phi$, corresponding to narrower region in the $a-b$ plane, as shown in Fig.~\ref{fig:sigma_ana}. Due to the similar reason, the coupling $\kappa$ needs to be at least $\sim 0.7$ in this case. 



\subsection{DM relic density and direct detection constraints}

In the NMSSM with almost decoupled singlet sector, the correct DM relic density can be achieved through the $\chi \chi \to \phi \phi$ annihilations in s-channel, t-channel, and u-channel. 
For t-channel and u-channel annihilation, only the coupling $V_{\phi \chi \chi}$ is relevant. As for s-channel annihilation, $V_{\phi\phi\phi}$ comes into play. 
Note that with negligible $V_{\phi\phi\phi}$, the density of SIDM tends to be under-abundant, because a large $\kappa$ and a light $\phi$ ($m_{\phi} < m_\chi$) is required for solving the small structure problems. 
Using the triple scalar interaction in NMSSM
\begin{align} 
V_{\phi\phi\phi}&=\frac{V_{SSS}}{2\sqrt{2}}=-i\sqrt{2}\kappa(3m_\chi+A_\kappa) \equiv -i \Lambda m_\phi~, \label{eq:vfff}
\end{align}
we can get the total DM annihilation cross section as
\begin{align}
&\sigma_{\text{ann}} v =d v^2+O(v^2)~,\\
&d=\frac{\kappa^2\sqrt{x^2-1}}{192\pi m_\phi^2}  \left(\frac{32\kappa^2x(9x^4-8x^2+2)}{(1-2x^2)^4} +\frac{3\Lambda^2}{x(1-4x^2)^2} -\frac{8\sqrt{2}\kappa\Lambda(5x^2-2)}{(1-2x^2)^2(4x^2-1)}\right)~,
\end{align}
where $x=m_\chi / m_\phi$.

\begin{figure}[thb]
	\begin{center}
		\includegraphics[width=0.8\textwidth]{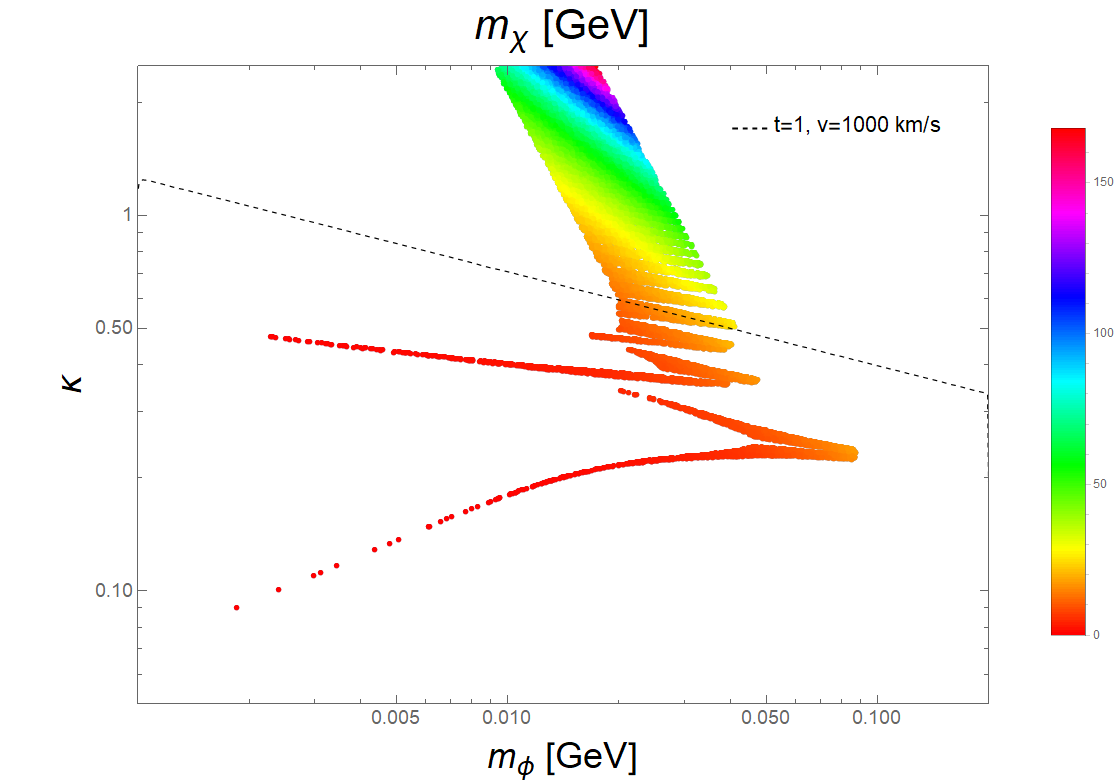}
	\end{center}
	\caption{\label{fig:sigrelic} Dots represent the region, where the small scale structure problem can be addressed, according to analytic formulas  $\sigma_{\text{V}}^{\text{Comb}}/m_\chi$. The value of $m_\chi$ of each selected point is fixed to make $\Omega h^2=0.12$ based on Eq.~\ref{eq:omega}. The dashed line satisfies $t=1$ when $v=1000$ km/s.}
\end{figure}

The relic density of DM can be estimated as~\cite{Lees:2014xha}
\begin{align} \label{eq:omega}
\Omega_\chi h^2=\frac{2\times 8.77\times10^{-11}\text{GeV}^{-2}x_f^2}{3 g_{*}^{1/2} d},
\end{align}
where $g_{*}\approx 10$ and $x_f\approx 10$.
It is noted that $d$ gets its minimum value (the relic density is maximal) as
\begin{align}
d_{min}=\frac{\kappa^2\sqrt{x^2-1}}{192\pi m_\phi^2}\left(\frac{64\kappa^2x(x^4-2x^2+1)}{3(1-2x^2)^4}\right)
\end{align}
when
\begin{align}
A_\kappa&=\frac{44x^4-16x^2-1}{3(1-2x^2)^2}m_\chi\approx \frac{11}{3}m_\chi~(x\gg 1).
\end{align}
Because the SIDM is under-abundant in most cases, this limit is useful to test whether the NMSSM with freeze-out mechanism can explain the current DM relic density. If the $\Omega_\chi h^2$ with $A_\kappa \sim (11/3) m_\chi$ is still less than $\sim 0.1$, either the model needs to be extended or other DM production mechanisms are required. 
In Fig.~\ref{fig:sigmaom}, the green lines correspond to the relations of $\kappa$ and $m_\chi$ such that the DM relic density calculated by Eq.~\ref{eq:omega} (taking  $A_\kappa = (11/3) m_\chi$) is equal to 0.12~\cite{Akrami:2018vks}. 
When $m_\phi$ is greater than 0.1 GeV, the green lines will not cross the blue regions and will further departure away as $m_\phi$ increases, as the small scale structure problem requires much larger $\kappa$ than the correct relic density does.
However, the green line crosses the blue regions when $m_\phi$ is 0.01 GeV and smaller, which means in this case, the small scale structure and the DM relic density can be addressed simultaneously. By fixing the value of $m_\chi$ to make $\Omega h^2=0.12$ according to Eq.~\ref{eq:omega}, Fig.~\ref{fig:sigrelic} shows the region where the small scale structure problem can be addressed on the $m_\phi-\kappa$ plane. There is an upper limit for $m_\phi$ ($\sim 0.09$ GeV). The selected points above (below) the dashed line ($t=1,~v=1000$ km/s) belong to the classical (quantum) regime on the cluster scales ($v\sim 1000$ km/s), and all the selected points are in the quantum regime on the dwarf scales.


Moreover, many DM underground direct detection experiments have put stringent limit on SUSY DM models. For the singlino-like DM in the NMSSM, the nucleon-DM scattering cross section is proportional to the mixings between the singlet scalar and the Higgs boson (the one observed at the LHC, which is $H_u$-like): 
\begin{align}
\theta_{h\phi} \sim \frac{\mathcal{M}^2_{S,23}}{|\mathcal{M}^2_{S,22} - \mathcal{M}^2_{S,33}|} \sim \frac{\mathcal{M}^2_{S,23}}{(125~\text{GeV})^2} \sim \lambda \times \frac{2 \mu_{\text{eff}} v_u - v_d(A_\lambda +2 \kappa s + \mu^\prime)}{(125~\text{GeV})^2}~.~ \label{eq:hsmix}
\end{align}
And the magnitude of the spin independent proton-DM scattering cross section can be estimated by the following equation~\cite{PhysRevLett.106.121805}:
\begin{align}
\sigma^{\text{SI}}_{p\chi} \sim (\frac{\lambda}{10^{-7}})^2  (\frac{\kappa}{0.1})^2 (\frac{0.01~\text{GeV}}{m_\phi})^4 \left[ \frac{\mu}{m_Z} (\frac{A_\lambda }{\mu \tan \beta} -1) + 0.184 \frac{v}{\mu} \right] ^2 \times 10^{-4}~\text{pb}~,~
\end{align}
where $\tan\beta =v_u/v_d$ and $v=\sqrt{v^2_u + v^2_d} = 174$ GeV. 
Since the SIDM scenario requests a relatively large $\kappa$ and light scalar mediator $\phi$, in order to suppress the nucleon-DM scattering cross section below $\sim 10^{-4}$ pb, an extremely small $\lambda$ is required, {\it i.e.} $\lambda \lesssim 10^{-7}$, unless the $A_\lambda$ is tuned appropriately to implement an exact cancellation. 
As the SIDM mass is typically $\mathcal{O}(10^{-1} - 10^1)$ GeV , the most stringent limits come from XENON1T~\cite{Aprile:2018dbl}, CRESST~\cite{Angloher:2015ewa} and DarkSide50~\cite{Agnes:2018ves} experiments.  

\subsection{Parameter scan}
The parameters in the NMSSM most relevant to the scalar sector and neutralino sector are
\begin{align}
\lambda,~\kappa,~\tan \beta, ~ \mu_{\text{eff}}, ~A_\lambda, ~A_\kappa, ~m_H^2, ~\mu^\prime, ~m_S^{\prime 2}, ~\xi_F, ~\xi_S~.~
\end{align}
We perform random scan of these parameters in the ranges:
\begin{align}
\tan \beta \in [1,50], ~\lambda \in [10^{-8},10^{-5}], ~ \kappa \in [0.05, 0.3],~\mu_{\text{eff}} \in [300,2000]~\text{GeV}, \nonumber \\ 
 \xi_F \in [-1,1] \times 10^{11}~\text{GeV}^2,~m^2_{H} \in [10^8,10^{12}]~\text{GeV}^2~.~ 
\end{align}
Note that the $\lambda$ is scanned logarithmically ({\it i.e.} $\lambda=10^{r}$ and $r \in [-8,-5]$ is a uniform random number), in order to focus on the small $\lambda$ region. 
To implement the SIDM scenario in the NMSSM, some of the parameters are scanned in much smaller regions:
\begin{align}
\mu^\prime &= (- 2 \kappa s) \pm 20~\text{GeV} ~,~ \label{eq:mchi}\\ 
\xi_S &=  \lambda v_u v_d (A_\lambda + \mu^\prime) +\kappa s^2 (A_\kappa + 4 \kappa s + 3 \mu^\prime) - \xi_F \mu^{\prime} - s \times [\exp(-5), \exp(-2) ] ~,~ \label{eq:mh1}\\
A_\lambda &= (2 \kappa s + \mu^\prime) \pm 40~\text{GeV}~.~ \label{eq:mixsh}
\end{align}
According to the discussions in Sec.~\ref{sec:nmssm}, Eq.~\ref{eq:mchi}, Eq.~\ref{eq:mh1} and Eq.~\ref{eq:mixsh} lead to light singlino DM, light CP-even singlet scalar, and small mixing between singlet scalar and doublet Higgs, respectively. 
Moreover, in order to suppress the DM annihilation $\chi \chi \to \phi \phi$, we scan
\begin{align}
A_\kappa  = (1.0 \pm 0.2) \times \frac{11(2 \kappa s + \mu^\prime)}{3}~.~
\end{align}
And $m_S^{\prime 2}$ is set to give heavy CP-odd singlet scalar, decoupling from the singlet sector. 

The rest of the NMSSM parameters are fixed as
$M_1=M_2 =3$ TeV, $M_3=5$ TeV, $m_{\tilde{L}_{1,2,3}} = m_{\tilde{E}_{1,2,3}} = m_{\tilde{Q}_{1,2,3}} = m_{\tilde{u}_{1,2,3}} = m_{\tilde{d}_{1,2,3}} =5$ TeV, $A_e=0$ GeV. Except that $A_t,A_b$ are scanned in the range of $[0,3]~\text{TeV}$ to give a SM-like Higgs mass around 125 GeV. 

The scanning is performed with the package NMSSMtools~\cite{Ellwanger:2004xm,Ellwanger:2005dv}, imposing the following phenomenological constraints as preselections:

\begin{itemize}
\item All LEP and Tevatron constraints that are implemented in the NMSSMtools. 
\item The second lightest CP-even scalar ($h_2$) being SM like, {\it i.e.} $m_h \in [120,130]$ GeV and coupling strength to SM particles in the range [0.9,1.1]. 
\item The lightest CP-even scalar $h_1$ and the lightest neutralino being singlet-like, {\it i.e.} with singlet component of the mixing matrix greater than 0.9. 
\end{itemize}

To implement a realistic SIDM scenario in the NMSSM, the parameter $\lambda$ as small as $\sim10^{-7}$ is required, such that the DM sector is isolated from the SM sector and the DM-nucleon cross section is suppressed. However, with $\lambda \lesssim 10^{-7}$ and $\mu_{\text{eff}} \gtrsim 100$ GeV (required by the chargino search at the LEP), the VEV of the singlet field $s = \mu_{\text{eff}}/ \lambda \sim 10^{9}$ GeV.  Moreover, the SIDM scenario favors light DM and large $\kappa$. According to Eq.~\ref{eq:mchi} and Eq.~\ref{eq:mh1}, $\mu^\prime \sim 10^{8}$ GeV and $\xi_S \sim 10^{25}$ $\text{GeV}^3$ (given $\kappa \sim \mathcal{O}(0.1)$). 
The light CP-even scalar (small $\mathcal{M}^2_{S,33}$) requires almost exact cancellation between numbers of order $10^{16}$. Although the parameter $\xi_S$ is set to implement this relation, it is numerically unstable, because we are using double precision floats, the machine precision of which are around $10^{-16}$. 
So in practice, we scan the $\mathcal{M}^2_{S,33}$ in the range $[e^{-5},e^{-2}]$, and use $\mathcal{M}^2_{S,33}=0$ to solve the $\xi_S/s$.

\begin{figure}[thb]
\begin{center}
\includegraphics[width=0.3\textwidth]{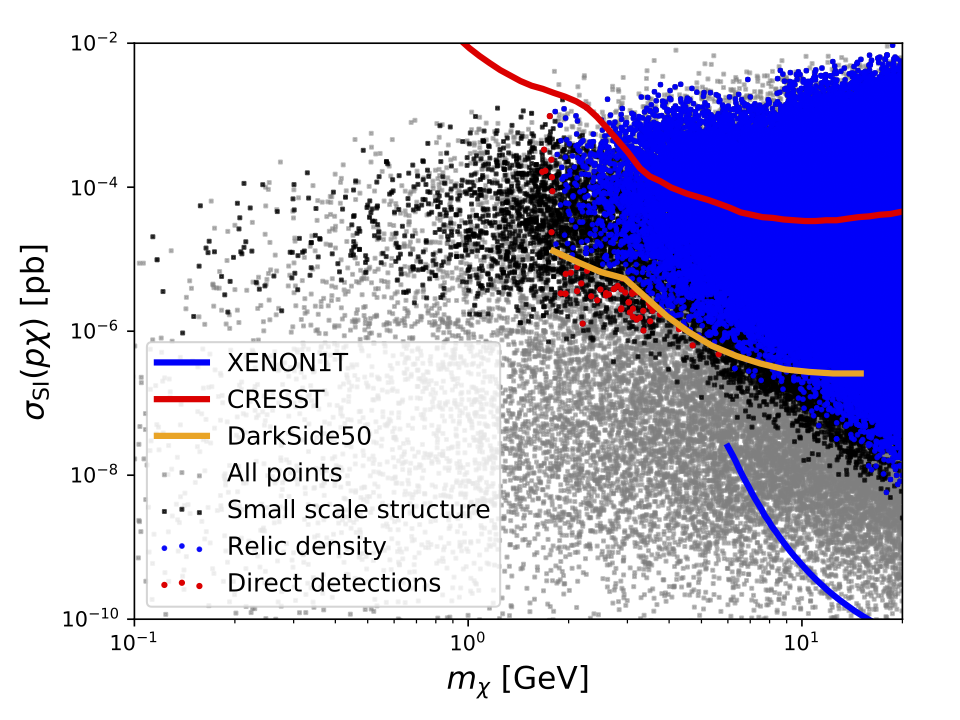}
\includegraphics[width=0.3\textwidth]{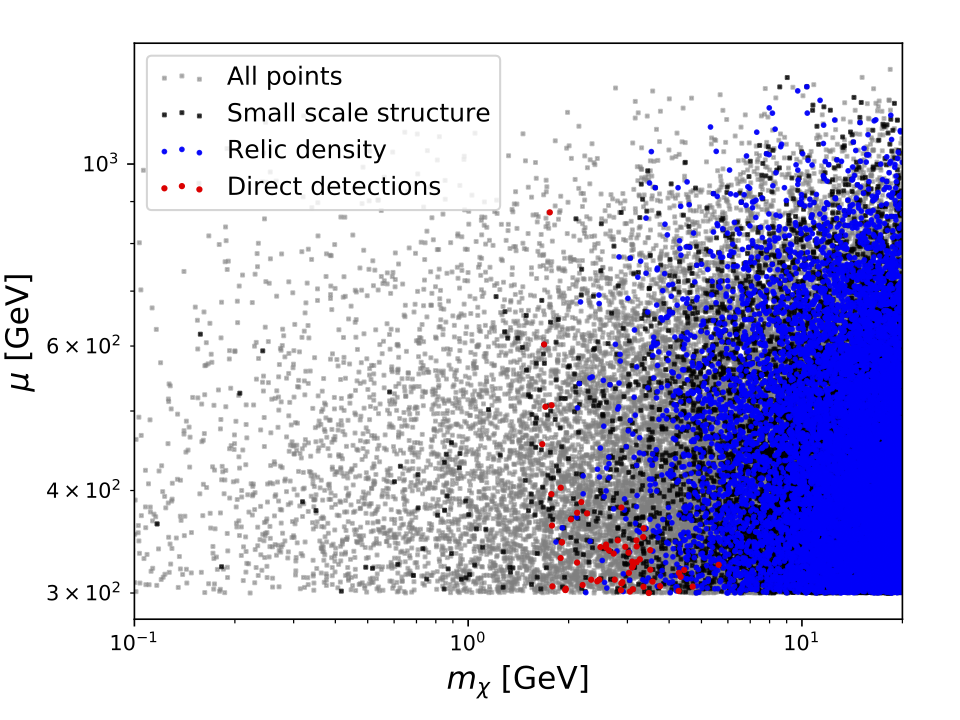}
\includegraphics[width=0.3\textwidth]{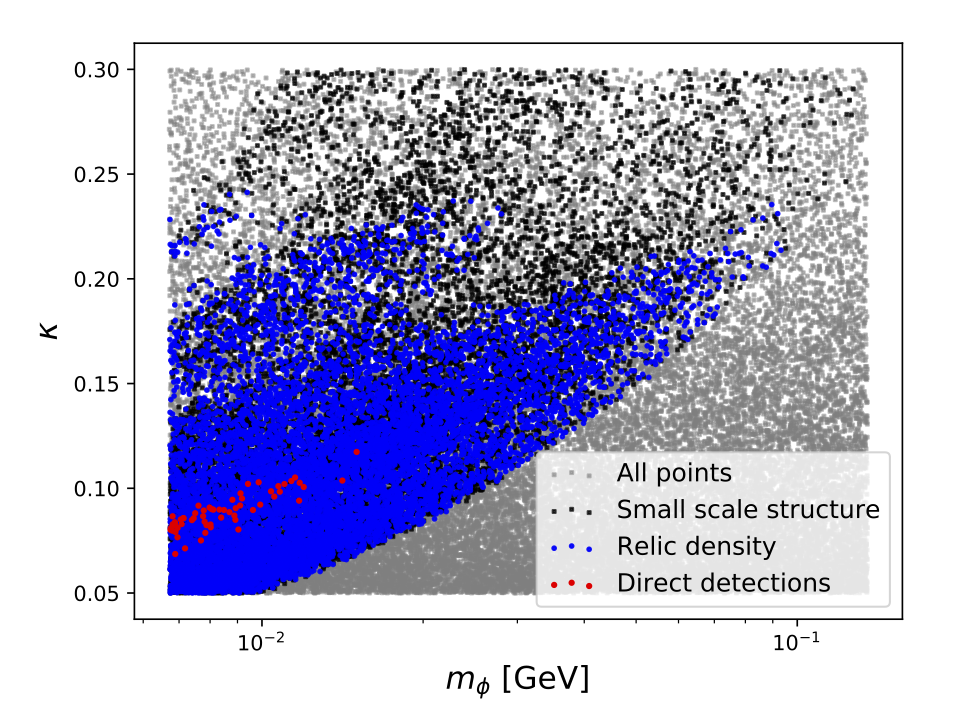}\\
\end{center}
\caption{\label{fig:nmssmscatter} The scatter plots of parameter space that pass the preselections (grey box). The small scale structure constraint (black box), DM relic density constraint (blue dot) and DM direct detection constraints (red dot) are applied in order. }
\end{figure}

The results of the parameter scan are demonstrated in Fig.~\ref{fig:nmssmscatter}, where the grey boxes correspond to the points that pass the preselections. Moreover, The small scale structure constraint ($\sigma^{v=10~\text{km/s}}_\text{V} /m_\chi \in [1,10]~\text{cm}^2$/g and $\sigma^{v=1000~\text{km/s}}_\text{V} /m_\chi \in [0.1,1]~\text{cm}^2$/g), DM relic density constraint ($\Omega h^2 \in [0.09,0.11]$) and DM direct detection constraints (XENON1T, CRESST and DarkSide50) are applied in order. Points that pass those constraints are shown with different colors. 
Due to the scanning range that we choose for $\lambda$, the typical spin independent cross section between DM and proton $\sigma_{\text{SI}} (p\chi)$ is $\sim 10^{-7}$-$10^{-3}$ pb. So the DM direct detection constraints are stringent. 
An even smaller $\lambda$ is still possible since it is not relevant for implementing the SIDM scenario. However, we will show later that smaller $\lambda$ renders long-lived singlet scalar, which will not provide any detectable signatures at colliders, except missing transverse momentum.  
The relic density constraint excludes all of the points with dark matter mass $m_\chi \lesssim 1$ GeV, because $\Omega h^2 \propto m_\chi^4$ according to Eq.~\ref{eq:omega}. 
Taking into account the direct detection constraints, the survival points have DM mass $\sim$ 1-5 GeV. 
It is interesting to observe in the middle panel of Fig.~\ref{fig:nmssmscatter} that the $\mu_{\text{eff}}$ ({\it i.e.} the Higgsino mass) is bounded from above, especially when the DM mass is heavier than 2 GeV. 
In order to address the small scale structure problem, the $\kappa/m_{\phi}$ needs to be large. However, relic density and DM direct detection constraints favor small value of $\kappa$. They put an upper limit on $m_{\phi}$, {\it i.e.} $m_{\phi} \lesssim 0.02$ GeV.

\begin{figure}[htb]
\begin{center}
\includegraphics[width=0.45\textwidth]{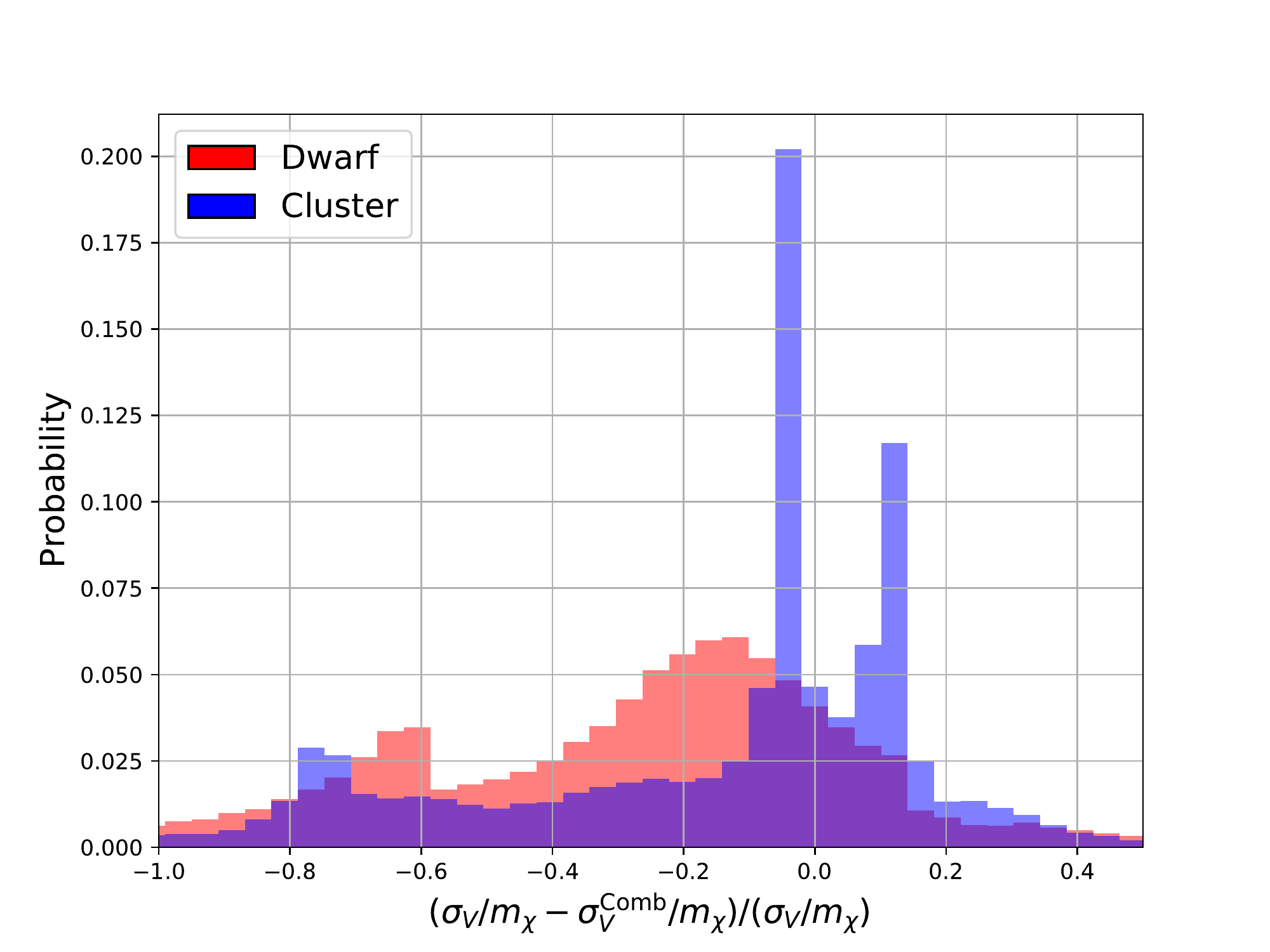} 
\includegraphics[width=0.45\textwidth]{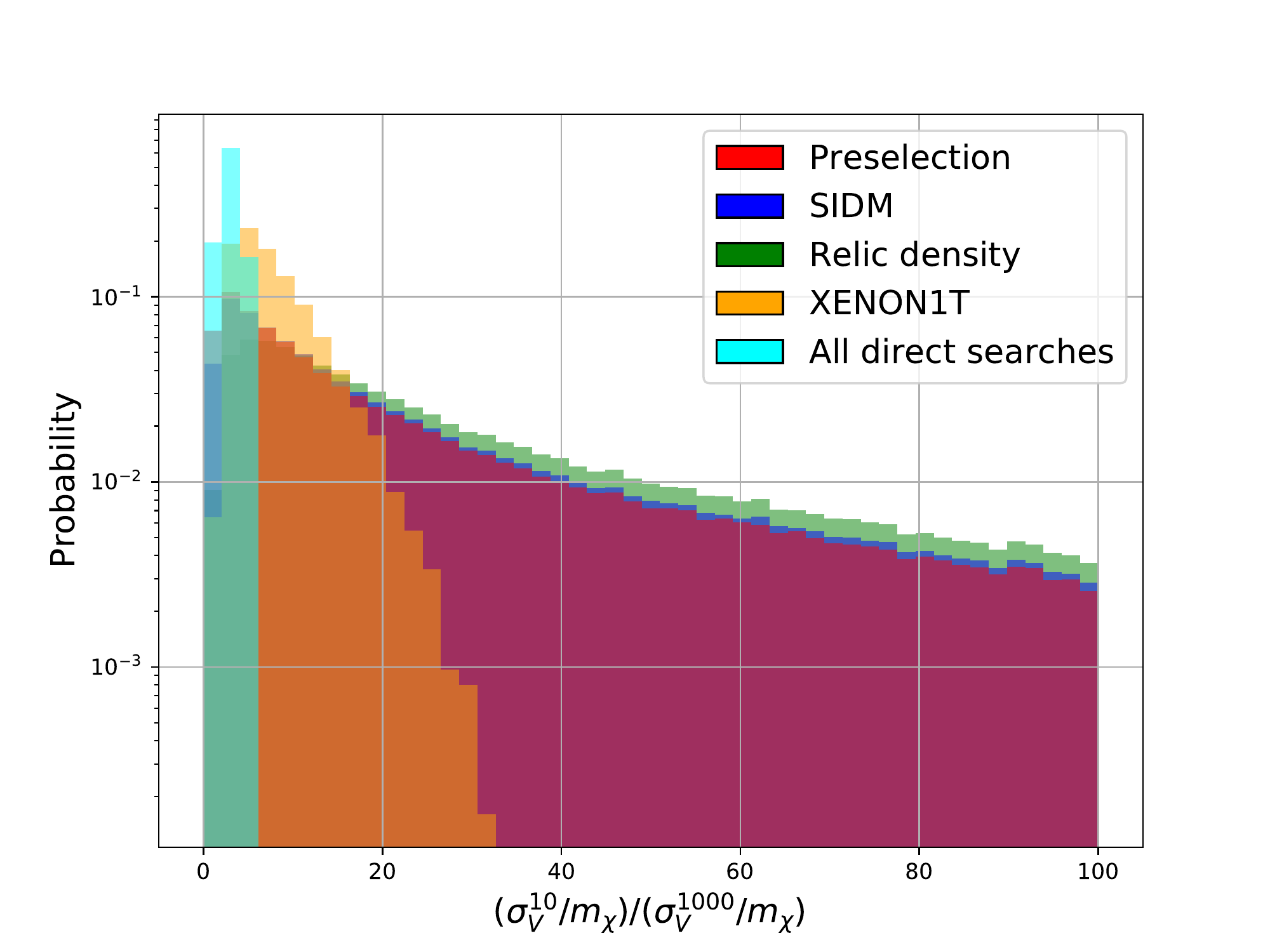}
\end{center}
\caption{\label{fig:compsigmav} Left panel: the relative difference between the analytical viscosity cross sections (divided by $m_\chi$) as given in Eq.~\ref{eq:sigma_comb} and numerically calculated ones (denoted by $\sigma_\text{V} /m_\chi$). The values for all preselected NMSSM points are used. Right panel: the distributions of the ratio between the $\sigma_\text{V} /m_\chi$ on dwarf and cluster scales  for scanned points in the NMSSM with different selection conditions.}
\end{figure}

The spin averaged viscosity cross sections for the points presented in Fig.~\ref{fig:nmssmscatter} are obtained by numerically solving the Schr{\" o}dinger equation with Yukawa potential~\cite{Tulin:2013teo}. In Sec.~\ref{sec:simp}, we have also presented analytical expression for the same variables. For comparison, in the left panel of Fig.~\ref{fig:compsigmav}, we show the distributions of the relative difference between the analytically and numerically calculated $\sigma_\text{V}/m_\chi$. 
There are more points with $\sigma_\text{V} /m_\chi < \sigma^{\text{Comb}}_\text{V} /m_\chi$ than the opposite case. The relative deviation between two methods is $\sim \mathcal{O} (10)\%$.
As discussed in Sec.~\ref{sec:simp}, the $\sigma_\text{V}$ can vary across several orders of magnitudes for different $a$ and $b$, such amount of deviation is acceptable for the purpose of estimation. 
The velocity dependence of the DM self-interacting cross section is illustrated in the right panel of Fig.~\ref{fig:compsigmav}. It can be found that the direct detection constraints are in contradiction with the velocity dependent feature. Parameter points that pass all DM direct detection constraints can only have the ratio between the $\sigma_\text{V} /m_\chi$ on dwarf and cluster scales at most around 3. 

\section{Splitting functions for singlet scalar and singlino}  \label{sec:splitting}

The SIDM scenario requires light scalar mediator and relatively large DM self-coupling. 
Such parameter setup may lead to DM/mediator showering: the particle can split multiple times during its propagation. 

Considering a time-like branching process $A\to B+C$ with the off-shell particle A being in the final state of a preceding hard process, the opening angle between $A$ and $B$ ($C$) is denoted by $\theta_b$ ($\theta_c$). 
In the high energy limit, the case that the daughters $B$ and $C$ are nearly collinear to the parent particle $A$ dominates the branching rather than the non-collinear case, as the scattering amplitude is proportional to 
\begin{align}
\frac{1}{(p^\mu_b+p^\mu_c)^2} \simeq \frac{1}{2 E_b E_c (1-\cos\theta)} = \frac{1}{Q^2}~.~
\end{align}
We parameterize the collinear time-like branching as follows. $\theta=\theta_b+\theta_c$ is the open angle and satisfies $\theta\ll 1$. The particle mass, energy, and momentum are $m_i$, $E_i$, and $P_i=|\vec{P}_i|$~($i=a,b,c$), respectively. $Q(>0)$ is the virtuality of $A$, which satisfies $Q^2=\left( E_a^2-P_a^2\right) \gg \left( E_{b,c}^2-P_{b,c}^2\right) $. Because $P_a\approx P_b+P_c$ in the collinear branching, we use $z=\frac{P_c}{P_a}$ and $\bar{z}=1-z$ representing the momentum fractions of $A$ taken up by $C$ and $B$, respectively. It is not hard to verify that $Q^2\approx P_b P_c \theta^2=P_a^2 z\bar{z}\theta^2$ in the collinear branching. 
     
For the hard process with a final state particle $A$ being the parent particle of a collinear time-like branching, the differential cross section can be expressed as
\begin{align}
d \sigma_{X, B C} \simeq d \sigma_{X, A} \times d \mathcal{P}_{A \rightarrow B+C}~,
\end{align}
where $d \mathcal{P}_{A \rightarrow B+C}$ is the differential splitting function for the $A\to B+C$. Using the parameters defined above, the splitting function can be expressed as 
\begin{align}
 \frac{d \mathcal{P}_{A \rightarrow B+C}}{dz~d\ln Q^2}\approx \frac{1}{N} \frac{1}{16 \pi^{2}}  \frac{Q^2}{\left(Q^{2}-m_{a}^{2}\right)^{2}} \overline{\left|M_{\text{split}}\right|^2}~,
\end{align}
where $N=2$ when $B$ and $C$ are identical particles and $N=1$ when $B$ and $C$ are different. The $\overline{\left| M_{\text{split}}\right|^2}$ is the spin-averaged matrix-element square for the $A\to B+C$ branching process, which can be computed from the amputated $A\to B+C$ Feynman diagram with on-shell polarization vectors. 

\begin{table}[htb]
\centering
\begin{tabular}{ccc}  
\hline
Process &  $\lambda_a(\lambda_b),~\lambda_c$ & ${\left| M_{\text{split}}\right|^2}$ \\
\hline 
$\phi \to \phi +\phi$ &  & $\Lambda^2 m_\phi^2$  \\
$\phi \to \chi +\chi$ &  $\lambda_b=\lambda_c$ & $2\kappa^2\big(Q^2-\frac{m_\chi^2}{z(1-z)} \big) $ \\
$\phi \to \chi +\chi$ &  $\lambda_b=-\lambda_c$ & 
$2\kappa^2 m_\chi^2 \frac{(1-2z)^2}{z(1-z)}$ \\
$\chi \to \phi +\chi$ & $\lambda_a=\lambda_c$ & 
$2\kappa^2 m_\chi^2 \frac{(2-z)^2}{1-z}$ \\
$\chi \to \phi +\chi$ & $\lambda_a=-\lambda_c$ & 
$2\kappa^2\big( Q^2z-\frac{m_\chi^2 z+m_\phi^2(1-z)}{1-z} \big) $ \\
\hline
\end{tabular}
\caption{\label{tab:splitf} The $\left| M_{\text{split}}\right|^2$  for each splitting in the singlet sector of the NMSSM. The fermion helicity is labelled by $\lambda$.}
\end{table}

In the singlet sector of the NMSSM, branching processes of the singlino/singlet scalar include $\chi\to \phi+\chi$, $\phi\to \chi+\chi$, and $\phi\to \phi+\phi$. The $\left| M_{\text{split}}\right|^2$  of these processes which are depend on the helicities of the fermions are summarized in Tab.~\ref{tab:splitf}, giving
\begin{align}
&\frac{d \mathcal{P}_{\phi \rightarrow \phi+\phi}}{dz~d\ln Q^2}\approx \frac{\Lambda^2}{32\pi^2}\frac{Q^2 m_\phi^2}{(Q^2-m_\phi^2)^2}~,\\
&\frac{d \mathcal{P}_{\phi \rightarrow \chi+\chi}}{dz~d\ln Q^2}\approx \frac{\alpha}{4\pi}\frac{Q^2}{Q^2-m_\phi^2}\left( 1-\frac{4m_\chi^2-m_\phi^2}{Q^2-m_\phi^2} \right) ~,\\
&\frac{d \mathcal{P}_{\chi \rightarrow \phi+\chi}}{dz~d\ln Q^2}\approx \frac{\alpha}{4\pi}\frac{Q^2}{Q^2-m_\chi^2}\left( z+\frac{4m_\chi^2-m_\phi^2}{Q^2-m_\chi^2} \right)~.
\end{align}


The evolution of the final-state radiation (FSR) is dominated by the splitting functions. For the possible time-like branching of a parent particle A, the famous Sudakov form factor
\begin{align}
\Delta_A(Q_{\text{max}};Q_0)=exp\left[-\sum_{BC}\int_{\ln Q_0^2}^{\ln Q_{\text{max}}^2}d\ln {Q}^2 \int_{z_{\text{min}}(Q)}^{z_{\text{max}}(Q)} dz ~\frac{d \mathcal{P}_{A \rightarrow B+C}\left( z,Q\right)  }{dz~d\ln {Q}^2}\right]~,
\end{align} 
describes $A$'s probability of evolving from $Q_{\text{max}}$ to $Q_0$ without branching, where the allowed $z$ range $(z_{\text{min}}(Q),~z_{\text{min}}(Q))$ at $Q$ depends on kinematics and is given by
\begin{align}
 &v_b  =\sqrt{1-\left( \frac{2 m_b Q}{{Q}^2+m_b^2-m_c^2}\right)^2}~,   \\
 &v_c  =\sqrt{1-\left( \frac{2 m_c Q}{{Q}^2+m_c^2-m_b^2}\right)^2}~,    \\
&z_{\text{min}}(Q) =\frac{1-v_c}{1+v_c / v_b}~, \\
&z_{\text{max}}(Q) =\frac{1+v_c}{1+ v_c / v_b}~. 
\end{align}

We study the evolution of the FSR by a numerical Monte Carlo method with Markov chain based on the Sudakov factors of $\chi$ and $\phi$. The evolution is operated by running from a high virtuality scale $Q_{\text{max}}$, chosen to be the CM-frame energy of the hard partonic process, down to a low scale $Q_{\text{min}}$ with small $Q$ steps. If the branching of $A\to B+C$ takes place at some $Q$, the evolution will be carried on with both the daughters $B$ and $C$. 
In Fig.~\ref{fig:sigmaom}, we plot the contours of the probability of the $\chi \to \phi+\chi$ branching taking place, which is computed as $P_{\chi \to \phi \chi}=1-\Delta_\chi(Q_{\text{max}};m_\chi+m_\phi) $, with $m_\phi=$0.01, 0.1, 1.0 GeV. The branching probability increases with $\kappa$ and can be above 0.7 when $\kappa\sim 2.5$ and $m_\phi=0.01$ GeV. Lower initial energy $Q_{\text{max}}$ leads to smaller branching probability, especially when $Q_{\text{max}}$ approaches  $m_\chi+m_\phi$, which is the threshold for the branching. 
When $m_\phi$ increases, the branching probability drops quickly because the factor of $1 / \left( Q^2-m_\chi^2\right)^2 $ in the splitting function $\frac{d \mathcal{P}_{\chi \rightarrow \phi+\chi}\left( z,Q\right)  }{dz~d\ln {Q}^2}$ decreases dramatically with $m_\phi$ when $Q$ approaches $Q_{\text{min}}=m_\chi+m_\phi$.

\section{Production of the singlet scalar at the LHC} \label{sec:simplhc}
The couplings between the MSSM sector and singlet sector are suppressed by the tiny $\lambda$. However, the particles in the singlet sector can still be produced through the decay of SUSY particles at the LHC. In particular, as shown in Fig.~\ref{fig:nmssmscatter}, the effective $\mu$ parameter is find to be $\lesssim 1$ TeV for $m_{\chi} \lesssim 10$ GeV (note that we scan the $\mu_{\text{eff}}$ in  the range [300,2000] GeV). Thus, relatively light Higgsinos are predicted. 
Assuming R-parity conservation, the light Higgsinos can be pair produced (in terms of $\tilde{H}^\pm \tilde{H}^\mp$, $\tilde{H}^\pm \chi_{2,3}$, $\chi_{2,3} \chi_{2,3}$ ) at the LHC through s-channel gauge boson exchange with the SM gauge couplings. 

At the tree level, the mass splitting between two neutral Higgsinos is $\sim m^2_Z/M_{1,2}$, and the splitting between charged Higgsino and the lighter neutral Higgsino is approximately half of that. Moreover, radiative corrections further induce $\sim \mathcal{O}(100)$ MeV mass difference between charged and neutral states~\cite{Drees:1996pk}.  
As a result, the heavier neutral Higgsino and the charged Higgsino will dominantly decay through $\chi_3 \to Z^*(\to ff) \chi_2$ and $\tilde{H}^\pm \to W^*(\to ff) \chi_2$, which typically happen in time scale of $\mathcal{O}(10^{-15})$ second (prompt decay inside the detector). 

The lightest Higgsino goes through two body decays: $\chi_2 \to Z \chi$, $\chi_2 \to H_2 \chi$ and $\chi_2 \to \phi \chi$ with branching ratio proportional to $\lambda^2$, $\lambda^2$ and $\kappa^2 \lambda^2$, respectively. Here the $H_2$ is the SM-like Higgs boson. 
With $\lambda \sim \mathcal{O}(10^{-7})$, the typical lifetime of the lightest Higgsino is $\mathcal{O}(10^{-10})$ second, so that it decays inside the detector. 
A relatively large $\kappa$ that is used to address the small scale structure problem, also leads to relatively large branching ratio of $\chi_2 \to \phi \chi$, which provide a unique opportunity to probe the singlet sector of the SIDM scenario in the NMSSM at colliders. 

\begin{figure}[htb]
\begin{center}
\includegraphics[width=0.45\textwidth]{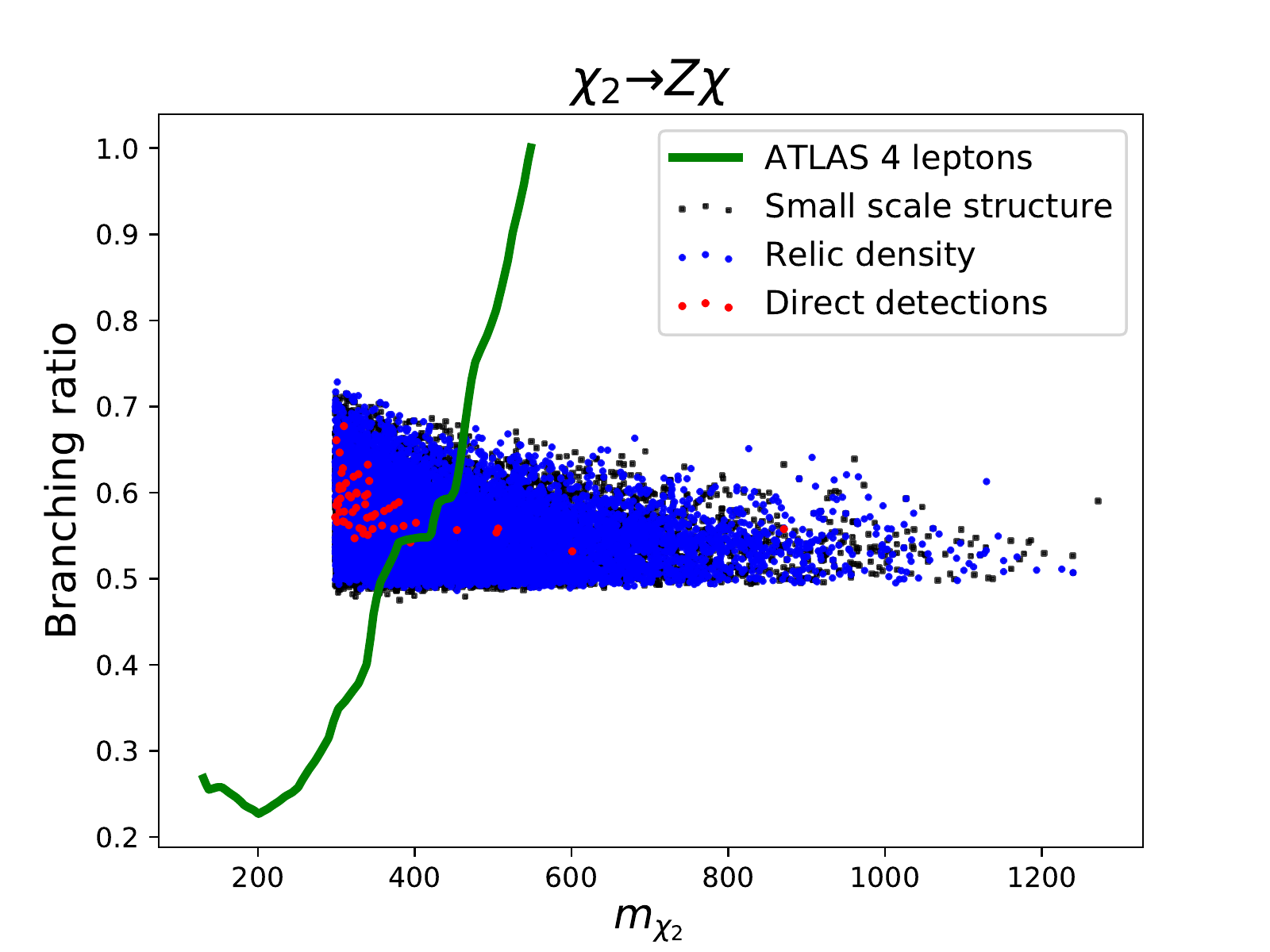}
\includegraphics[width=0.45\textwidth]{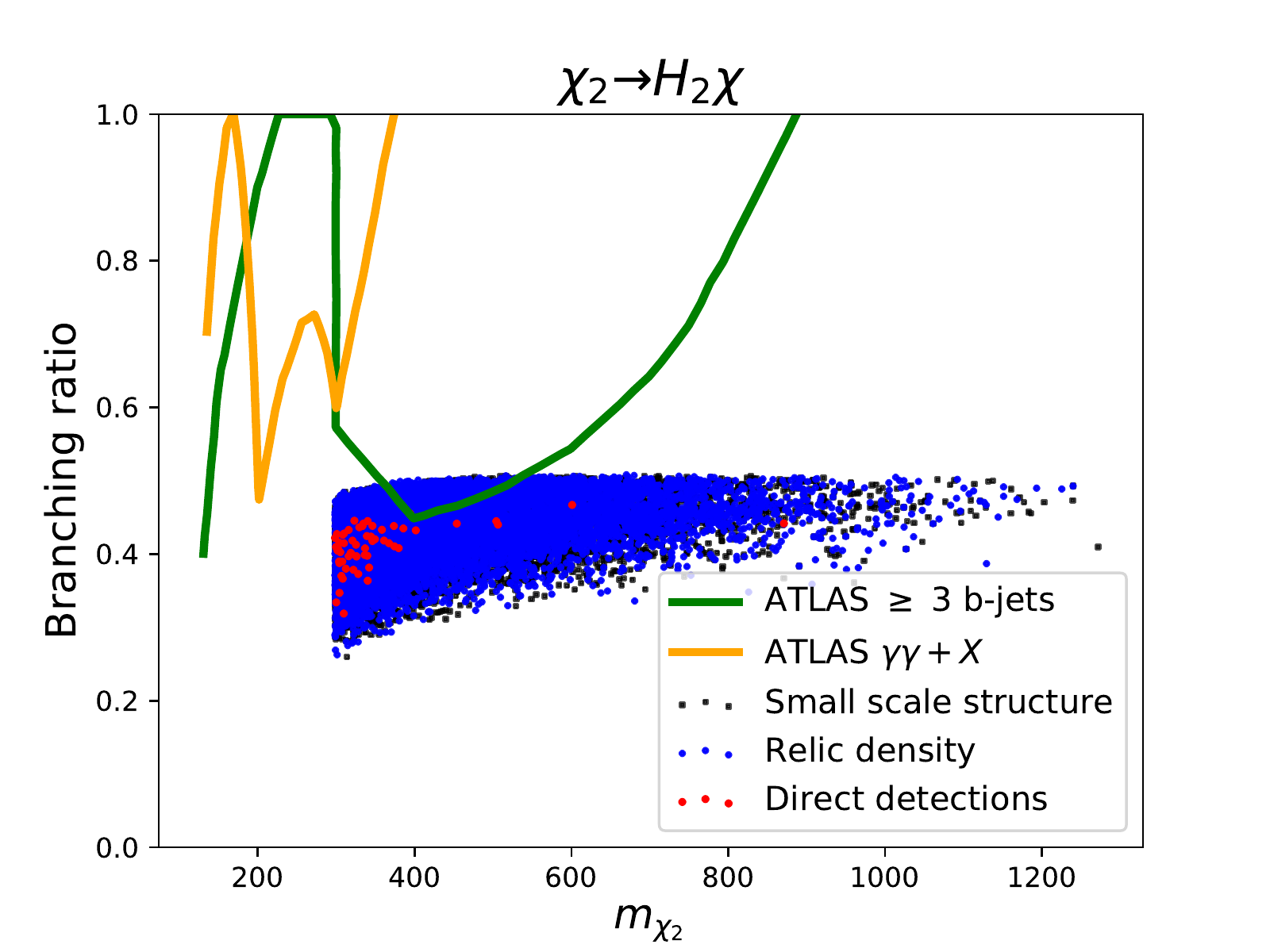}\\
\includegraphics[width=0.45\textwidth]{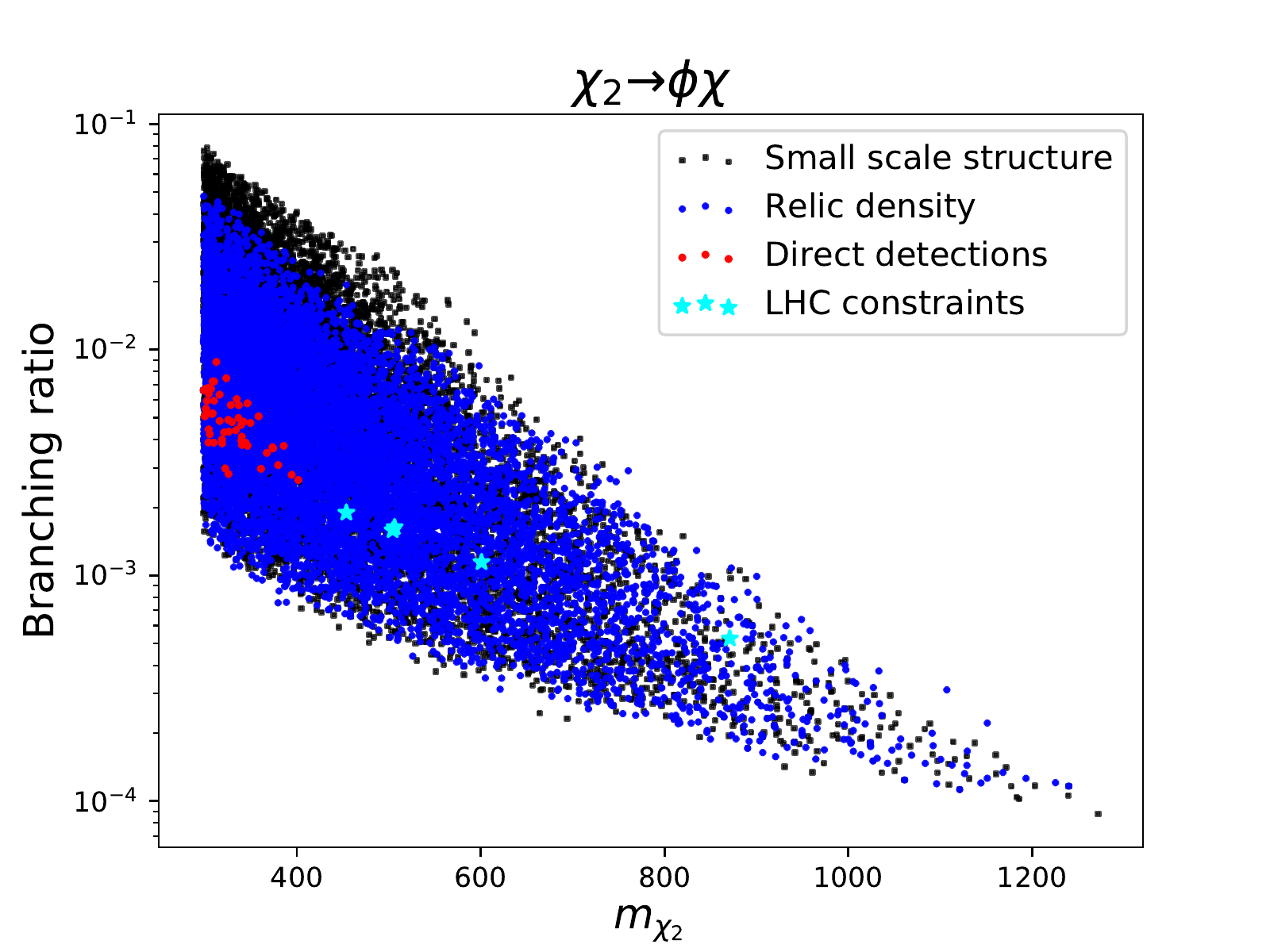}
\includegraphics[width=0.45\textwidth]{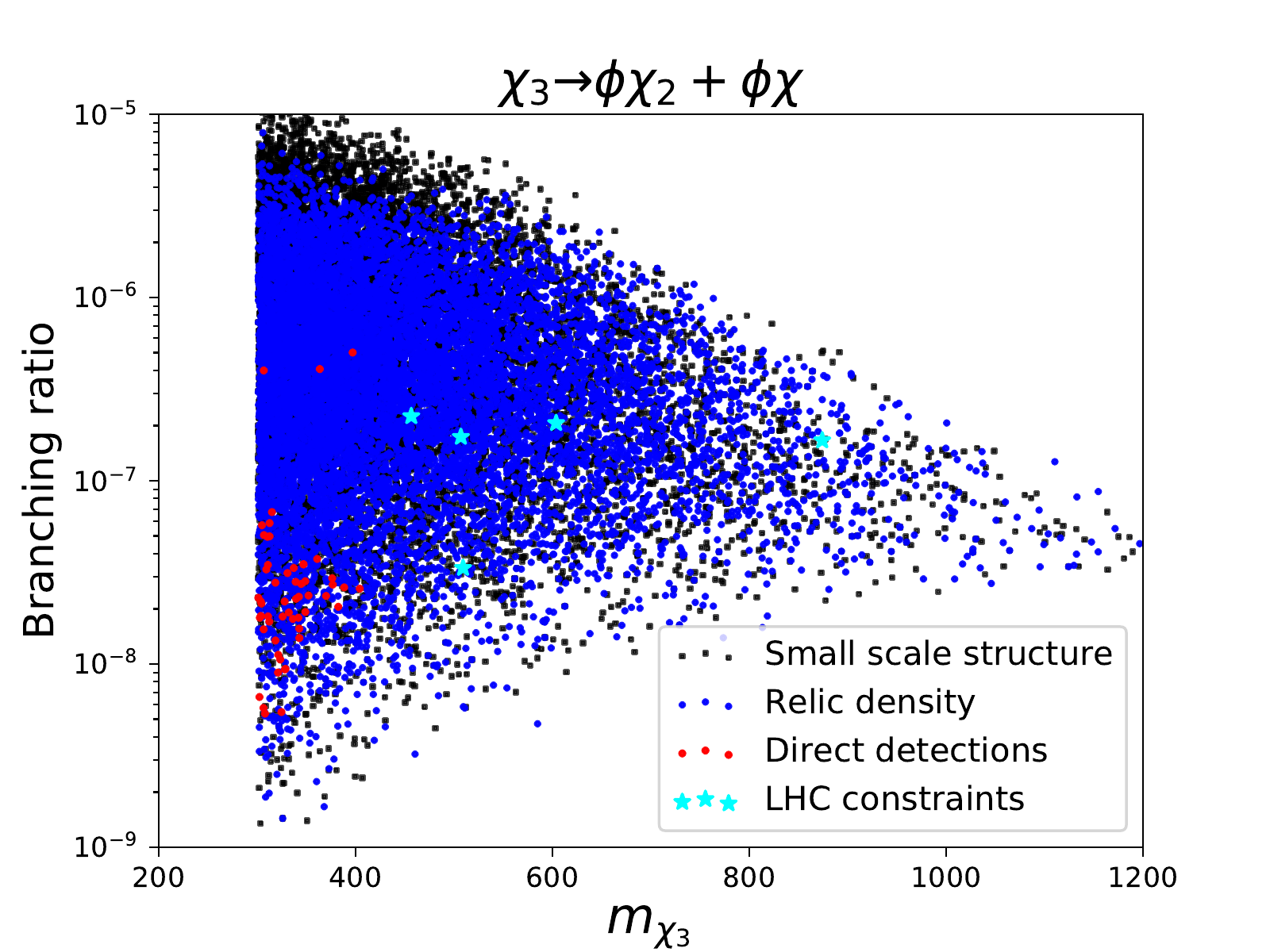} 
\end{center}
\caption{\label{fig:higgsinodecay} Decay branching ratios of the neutral Higgsinos ($\chi_2$ and $\chi_3$). 
Points pass different constraints are indicated in the same way as in Fig.~\ref{fig:nmssmscatter}. In the upper panels, the LHC search constraints~\cite{ATLAS:2020fdg,Aaboud:2018htj,Aad:2020qnn} on the Higgsino decaying into $Z$ boson and Higgs boson are imposed. In the lower panel, the points that also evade the LHC search constraints are marked in green. 
In the lower right panel, the summed branching ratios of $\chi_3 \to \phi \chi$ and  $\chi_3 \to \phi \chi_2$ are shown. }
\end{figure}

In Fig.~\ref{fig:higgsinodecay}, we present the decay branching ratios of the neutral Higgsinos for the scanned points. 
As before, the small scale structure constraints, the DM relic density constraint and the DM direct detection constraints are applied in order. 
For $\kappa \lesssim 0.1$, the lightest Higgsino will dominantly decay into $Z/H_2 + \chi$.
As the ATLAS collaboration has conducted searches for Higgsinos decaying into $Z$ boson and Higgs boson, some of the points can be excluded already. In upper panels, the constraints from ATLAS searches for four or more charged leptons~\cite{ATLAS:2020fdg}, at least three b-tagged jets~\cite{Aaboud:2018htj} and two photon + missing transverse momentum~\cite{Aad:2020qnn} are projected on. We can find the ATLAS search for four or more charged leptons excludes the points with Higgsino mass $\lesssim 400$ GeV, while the constraints on $\chi_2 \to H_2 \chi$ channel are much milder. 
Points that evade the LHC constraints and all others are shown by green dots in lower plots. The Higgsino masses for allowed points are $\sim 400-1000$ GeV and the corresponding production cross sections~\cite{Fuks:2012qx,Fuks:2013vua} are $\sim 88.7-0.969$ fb (summed cross sections of $\tilde{H}^\pm \tilde{H}^\mp$, $\tilde{H}^\pm \chi_{2,3}$, $\chi_{2,3} \chi_{2,3}$ productions at the 13 TeV LHC). 
The branching ratios of $\chi_2 \to \phi \chi$ for allowed points are $\sim \mathcal{O}(10^{-3})$. So, at the high-luminosity LHC, the signal rate of the Higgsino production with subsequent decay $\chi_2 \to \phi \chi$ in the SIDM scenario of NMSSM is sizable. 
In the lower right panel, we show the the summed branching ratios of $\chi_3 \to \phi \chi$ and  $\chi_3 \to \phi \chi_2$ for the heavier neutral Higgsino ($\chi_3$). As have been discussed above, the $\chi_3$ is dominated by the three body decay with off-shell gauge boson. Its branching ratio to the singlet-like scalar is highly suppressed by the $\lambda$ coupling. 

\subsection{Benchmark points}

For illustration purpose, we provide two benchmark points for the SIDM scenario of NMSSM in Tab~\ref{tab:benchmarknmssm}. With light mediator and relatively large $\kappa$, the spin averaged viscosity cross sections for both points are $ \gtrsim 1.0$  cm$^2$/g on dwarf scales, thus address the small scale structure problem.  
The production rates of the Higgsinos pair at the 13 TeV LHC with one of the Higgsino decaying through $\chi_2 \to \phi \chi$ for  benchmark point ($i$) and ($ii$) are 0.108 fb and 8.95 fb, respectively. 
The sizes are considerable and the corresponding signals are worth dedicated searches. Given the decay width of the lighter neutral Higgsino $10^{-14}$ GeV, it can travel a distance of $\sim \mathcal{O}(1)$ centimeter inside a detector before its decay. 

\begin{table}[h]
\begin{center} 
\resizebox{0.9\textwidth}{!}{
\begin{tabular}{|c|c|c|c|c|c|c|c|c|c|}
\hline
Benchmark & $m_{\tilde{\chi}_1}$[GeV] & $m_{h_1}$[GeV] & $\kappa$ & $A_\kappa$[GeV] & $\mu$[GeV] & $\lambda$ & $\mu^\prime$[GeV] & $\Omega h^2$   & $\sigma_{\chi p}^{SI}$[pb] \\\hline
({\it i}) &  -1.70  & 0.0068 & 0.083 & -6.78 & 505.9 & $5.80 \times 10^{-8}$ & $-1.45 \times 10^{9}$ & 0.098 & $1.69 \times 10^{-4}$   \\\hline
({\it ii}) & 20.0 & 0.023 & 0.237 & 77.05 & 363.0 & $7.57 \times 10^{-8}$ & $-2.27\times 10^{9}$ & 0.11 & $2.04 \times 10^{-5}$ \\
\hline
\end{tabular} }\\
\resizebox{1.0\textwidth}{!}{
\begin{tabular}{|c|c|c|c|c|c|c|c|c|}
\hline
 $m_{\tilde{\chi}_2}$[GeV] & $\Gamma(\tilde{\chi}_2)$[GeV] & $m_{\tilde{\chi}_3}$ & $\chi_2 \to H_1 \chi_1$ & $\chi_2 \to H_2 \chi_1$ & $\chi_2 \to Z \chi_1$ & $\chi_3 \to H_1 \chi_{1,2}$ & $\sigma^{v=10~\text{km/s}}_{\text{V}} /m$  &  $\sigma^{v=1000~\text{km/s}}_{\text{V}} /m$ \\\hline
504.2 & $ 1.58 \times 10^{-14}$ & 507.0 & 0.0016 & 0.445 & 0.554 & $1.74 \times 10^{-7}$ &  1.15 cm$^2$/g&  0.615 cm$^2$/g \\\hline
361.4 & $ 1.88 \times 10^{-14}$ & 364.1 & 0.033 & 0.464 & 0.503  & $2.74\times 10^{-6}$ & 2.09 cm$^2$/g&  0.261 cm$^2$/g \\ \hline
\end{tabular} } \end{center}
\caption{\label{tab:benchmarknmssm} Two benchmark points in the SIDM scenario of the NMSSM. Both points pass the preselections, address the small scale structure problem and provide correct DM relic density. The benchmark point ($i$) also evade all current DM direct searches while the second one is not. } 
\end{table}

The most remarkable collider signature of those two benchmark points is the showers of the singlino and the singlet scalar, which are produced by the Higgsino decays, leading to multiple scalars/DMs in the final state. 
The $\kappa$ is bounded from above due to the DM relic density constraints. So the DM splitting, as well as the scalar splitting into DM pairs is suppressed. According to our simulation, this kind of splittings happens less than twice per thousand events for our benchmark points. 
On the other hand, the probability of $\phi \to \phi \phi$ splitting can be sizable, since the coupling $V_{\phi \phi \phi}$ can be large (as given in Eq.~\ref{eq:vfff}).

\begin{figure}[htb]
\begin{center}
\includegraphics[width=0.45\textwidth]{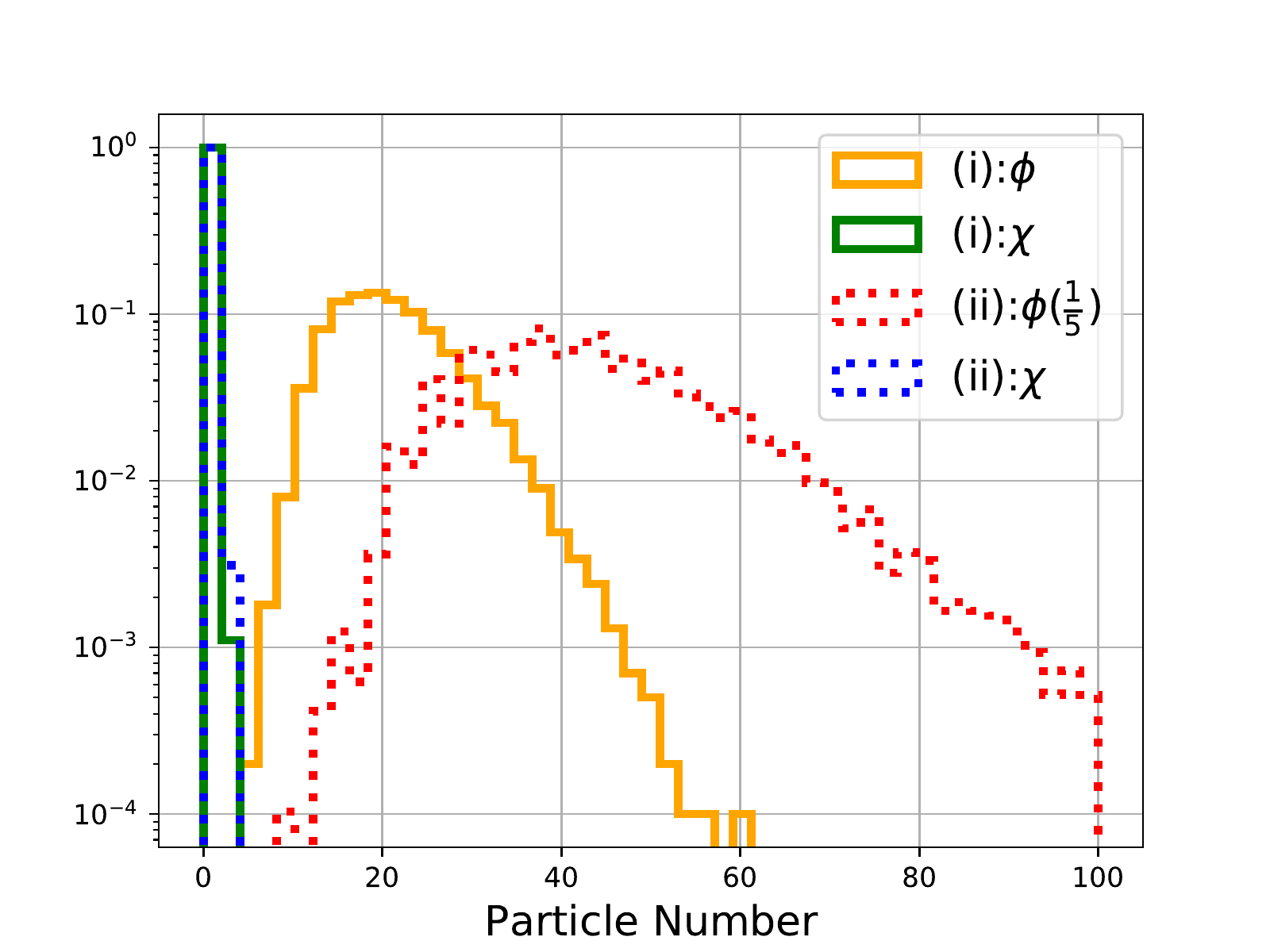}
\includegraphics[width=0.45\textwidth]{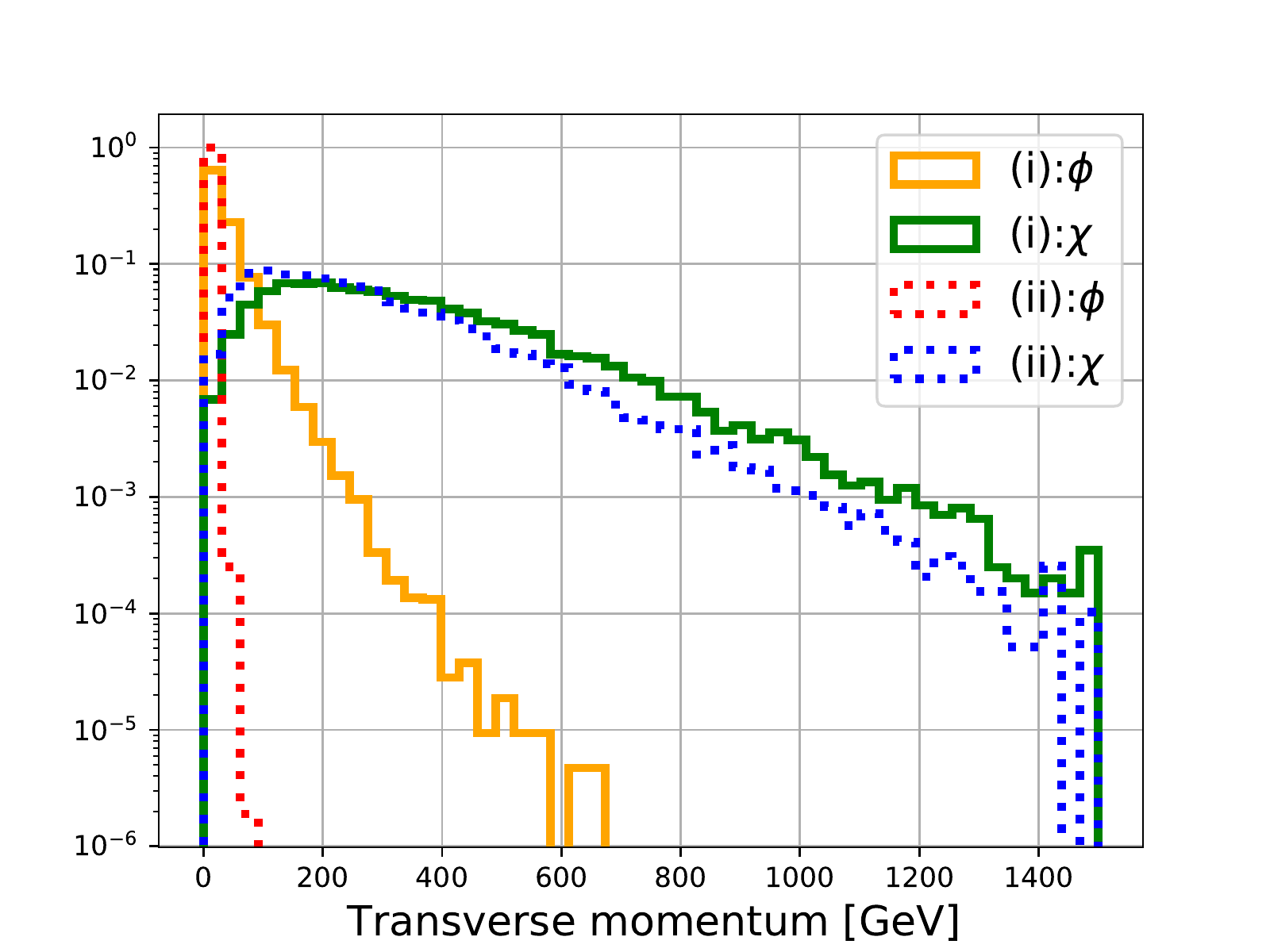}
\end{center}
\caption{\label{fig:nmssmbench} The multiplicities (left panel) and $p_T$ spectra (right panel) of the singlet scalar $\phi$ (singlino $\chi$) in the Higgsino production and decay process for benchmark point (i) and (ii). }
\end{figure}

In the left panel of Fig.~\ref{fig:nmssmbench}, we present the multiplicity of $\phi$ and $\chi$ from the process
\begin{align}
 p p \to \chi_2 \chi_2 \to (\phi \chi) (\phi \chi ) \label{eq:higproc}
\end{align}
with subsequent showers for our two benchmark points. 
Note the lightest Higgsino $\chi_2$ can be produced either from its direct production or from the decay of heavier Higgsinos. 
The final state $\phi$ particles have two origins: from the $\chi$ (denoted by $\chi \to \phi$) and the $\phi$ (denoted by $\phi \to \phi$) splittings in the process~\ref{eq:higproc}. 
For both benchmark points, the final state $\phi$ particles are mostly from the $\phi$ splitting. As the coupling $|V_{\phi \phi \phi}|$ are 1.40 GeV and 45.9 GeV  for benchmark point ($i$) and ($ii$), respectively. The benchmark point ($ii$) has much higher $\phi$ multiplicity than the benchmark point ($i$) due to its larger $|V_{\phi \phi \phi}|$. 
Although it is difficult to obtain a precise analytic expression for the $\phi$ multiplicity distribution, the rough estimation for the order of the average multiplicity of $\phi$ can be obtained by  
\begin{align}\label{eq:ppp}
\bar{n}_{\phi}\approx \bar{n}_{\phi;\phi \to \phi} \sim \int_{\ln \left( 4m_\phi^2\right) }^{\ln Q_{\text{max}}^2}d\ln {Q}^2 \int_{z_{\text{min}}(Q)}^{z_{\text{max}}(Q)} dz ~\frac{d \mathcal{P}_{\phi \rightarrow \phi+\phi}\left( z,Q\right)  }{dz~d\ln {Q}^2}~,
\end{align}
where $\bar{n}_{\phi;\phi \to \phi}$ stands for the multiplicity of $\phi$ coming from the branching of $\phi$. When taking $Q_{\text{max}}=500$ GeV, the Eq.~\ref{eq:ppp} gives $\bar{n}_{\phi}\sim$ 28 and 2642 for the benchmark point ($i$) and ($ii$), respectively.   
In the right panel of Fig.~\ref{fig:nmssmbench}, the spectra of the final state $\chi$ and $\phi$ for two benchmark points are shown. 
The $p_T$ spectra of $\chi$ are harder than that of $\phi$, since the $\phi$ goes through multiple splittings while $\chi$ does not split. 
Due to the same reason, the $p_T(\phi)$ spectrum of benchmark point ($ii$) is much softer than that of benchmark point ($i$).  

\subsection{Beyond the NMSSM}
The $\kappa$ coupling can be as large as $\mathcal{O}(1)$ when addressing the small scale structure problem. 
However, in the NMSSM, in order to implement the correct relic density with freeze-out mechanism, we find $\kappa \lesssim 0.2$ is required (as shown in the right panel of Fig.~\ref{fig:nmssmscatter}). Moreover, applying all phenomenological constraints leads to sizable $V_{\phi \phi \phi}$ coupling. 
Thus the $\phi$ splitting is copious while the $\chi$ splitting is highly suppressed in the NMSSM. 

In this subsection, we try to study the splitting of singlet/singlino in a simplified model framework 
where the small scale structure problem can be addressed while leaving out the 
DM relic density and direct detection constraints (this may be realized in a UV-complete model, where the DM is not produced from the thermal freeze-out mechanism). 
As a result, the $\kappa$ can be large and the coupling of $V_{\phi \phi \phi}$ is a free parameter. 
For illustration, we consider two benchmark points:
\begin{itemize}
\item Benchmark ({\it iii}): $m_{\phi}=0.1$ GeV, $m_{\chi}=5.65$ GeV, $\kappa=2.5$, $A_\kappa=11/3 m_{\chi}$. 
\item Benchmark ({\it iv}): $m_{\phi}=0.1$ GeV, $m_{\chi}=5.65$ GeV, $\kappa=2.5$, $A_\kappa=-3 m_{\chi}$.
\end{itemize}
Since the DM self-interacting viscosity cross section is irrelevant to the $A_\kappa$ parameter, both points have $\sigma^{v=10~\text{km/s}}_{\text{V}} /m_\chi  =6.16$ cm$^2$/g and  $\sigma^{v=1000~\text{km/s}}_{\text{V}} /m_\chi \sim 0.938$ cm$^2$/g, addressing the small scale structure problem. 
Given $\kappa \sim 2.5$, the branching ratio of the $\chi_2 \to \phi \chi$ channel is dominant, since its partial width is proportional to $\kappa^2$. 
So the production rate for the $\phi$ and $\chi$ is much enhanced. Moreover, with sizable $\kappa$, the $\chi \to \phi \chi$ and $\phi \to \chi \chi$ splittings are no longer negligible. 

\begin{figure}[htb]
\begin{center}
\includegraphics[width=0.45\textwidth]{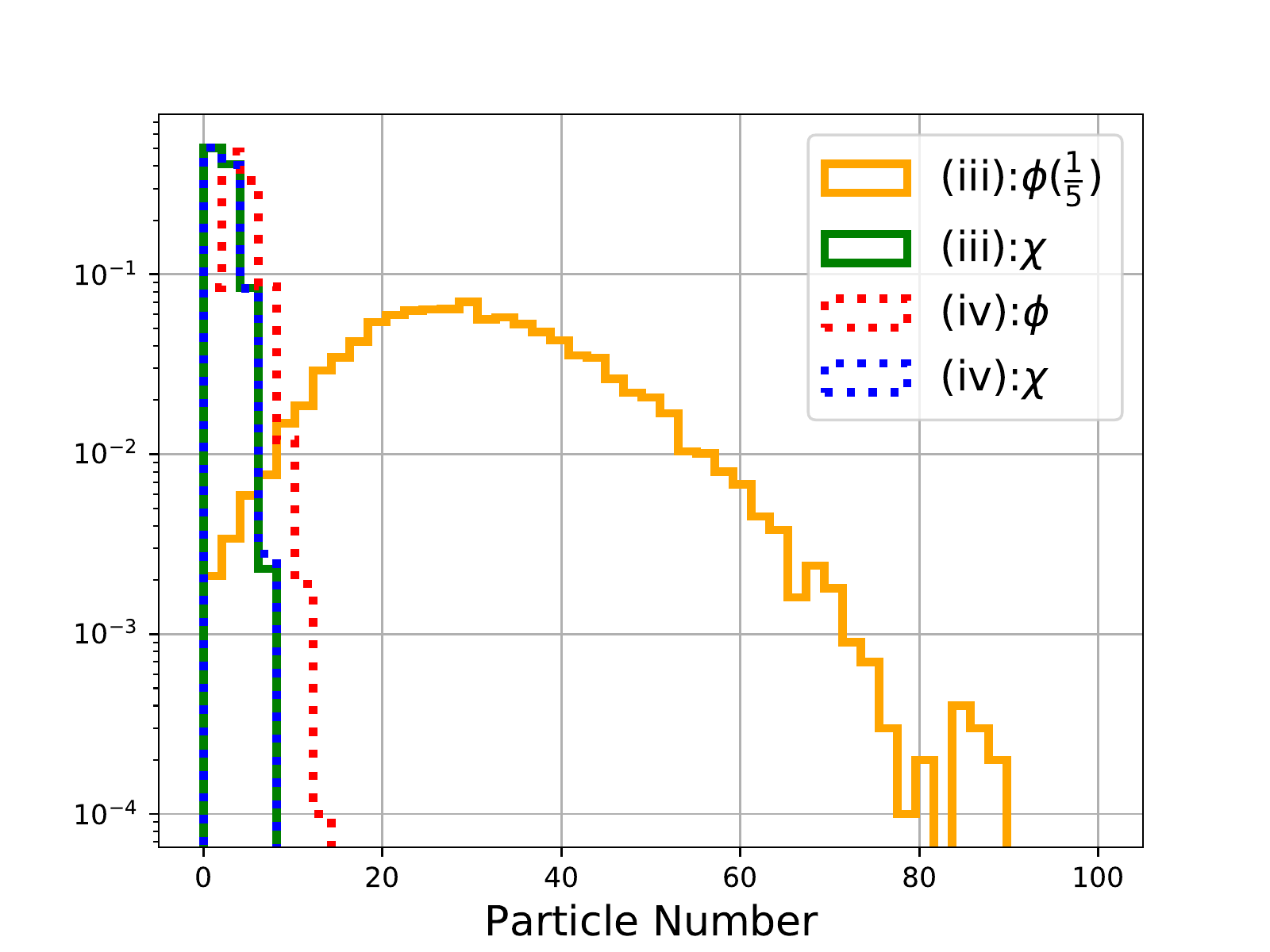}
\includegraphics[width=0.45\textwidth]{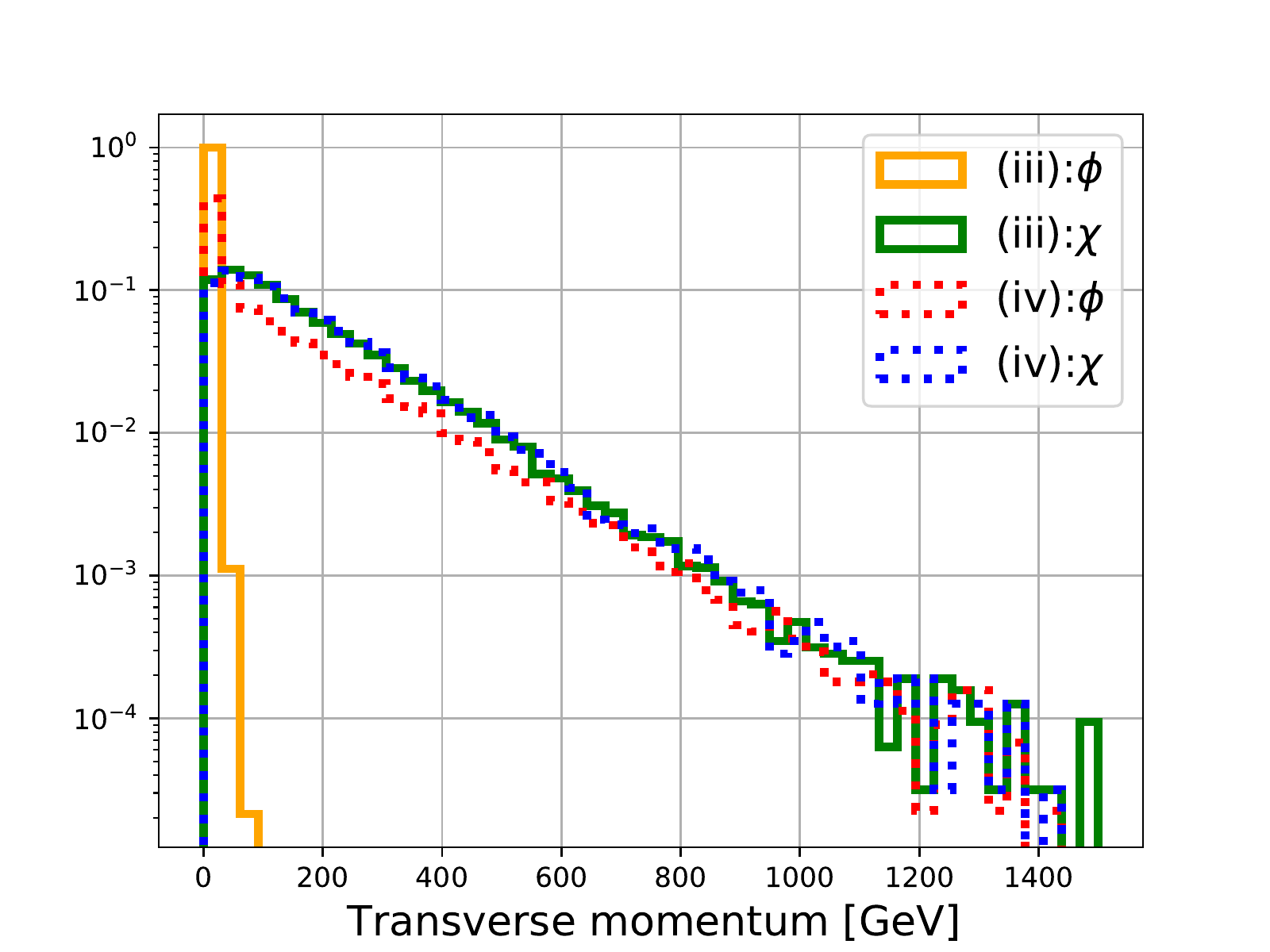}
\end{center}
\caption{\label{fig:benchmark34}The multiplicities (left panel) and $p_T$ spectra (right panel) of the singlet scalar $\phi$ and the singlino $\chi$ in the Higgsino production and decay process for benchmark point ($iii$) and ($iv$). The lightest Higgsino mass is taken to be 350 GeV.}
\end{figure}

In Fig.~\ref{fig:benchmark34}, we present the multiplicities and $p_T$ spectra for the singlet scalar $\phi$ and the singlino $\chi$. Similar as before, we are considering the process in Eq.~\ref{eq:higproc} for their production, and the mass of the lightest Higgsino is set to 350 GeV. 
Because the $|V_{\phi \phi \phi}|$ coupling is 133 GeV for benchmark point ($iii$) and vanishes for benchmark point ($iv$), we can expect that the $\phi$ multiplicity from $\phi \to \phi$ of benchmark point ($iii$) is high. 
There is also a certain amount of $\phi$ coming from the $\chi$ splitting, due to the large $\kappa$. So the $\phi$ multiplicity of benchmark point ($iv$) can also reach $\sim 5$ in spite of vanishing $V_{\phi \phi \phi}$.  
Two benchmark points have similar distributions of $\chi$ multiplicity, and the $\chi$ particles is less abundant comparing to the $\phi$ particles. 
As for the $p_T$ spectrum, the $\phi$ of the benchmark point ($iii$) is much softer than $\phi$ of the benchmark point ($iv$) and $\chi$ in both points, mainly because of the high multiplicity of $\phi$ such that the energy of the original $\phi$ is shared among its daughters.

\subsection{Decay of the singlet scalar}

The light scalar meditator in SIDM scenarios usually has mass less than twice of the pion mass, {\it i.e.} $m_\phi \lesssim 270$ MeV. In this case, it mainly decays into muons and electrons.
The leading order decay width is given by 
\begin{equation}
{\Gamma}_{\phi \ell \ell}=\theta_{h\phi}^2 \frac{m_\phi}{8\pi v^2} m_\ell^2 \beta^3~,
\end{equation}
where $\theta_{h\phi}$ is the mixing angle in the scalar sector (given by Eq.~\ref{eq:hsmix} in NMSSM), and $\beta=\sqrt{1-\frac{4m_\ell^2}{m_\phi^2}}$.  
With $\theta_{h\phi} \propto \lambda \sim 10^{-7}$, the typical lifetime of the singlet scalar (with mass $\sim 10$ MeV) can reach $\mathcal{O}(10^{4})$ seconds. 
As a result, the Higgsino decay and the subsequent showering inside the detector can only produce invisible final states. 
Moreover, the existence of a light long-lived scalar may spoil the success of the Big Bang Nucleosynthesis (BBN). 

In fact, the tension between the BBN (which requires lifetime of the mediator $\tau_\phi \lesssim 1$ second) and direct detection constraints (which favor small mediator-SM Higgs mixing) is commonly exist in SIDM scenarios with scalar mediator. There have been many solutions proposed to alleviate the tension.
One class of solutions assumes different DM thermal history~\cite{Baldes:2017gzu,Duch:2019vjg,Zhu:2021pad}. For example, in Ref.~\cite{Duch:2019vjg}, they assume the scalar mediator to be stable and annihilate into an extra scalar, constituting a subdominant DM. 
In the second class of solutions, the DM direct detection rate is suppressed by introducing CP-violating scalar sector~\cite{Kahlhoefer:2017umn} or inelastic DM-nucleon scattering~\cite{Blennow:2016gde}, so that a large singlet scalar-SM Higgs mixing is allowed, {\it i.e.} short lifetime of the scalar. 
And the simplest solution is to introduce new particles and interactions~\cite{Kainulainen:2015sva,Kang:2016xrm,Barman:2018pez} such that the scalar mediator can decay almost promptly through new channels. 

Different solutions will lead to dramatically different phenomena.  
For the solutions in later two classes, the scalar mediator can decay into visible final states inside the detector. Each $\chi$ or $\phi$ from the Higgsino decay will lead to a jet-like object with the identities of its constituents depending on the decay modes of the scalar mediator. 
Taking the $\phi \to e^+ e^-$ as an example (which is the dominant decay channel for our benchmark points), the corresponding signatures of $\chi$ and $\phi$ from Higgsino decay for each benchmark point are listed in Tab.~\ref{tab:signa}. 
Due to the small $\kappa$ of benchmark point ($i$) and ($ii$), the $\chi$ does not split and simply behaves as missing transverse energy ($\slashed{E}_T$) at the detector. 
After the shower, the $\phi$ for benchmark point ($i$) will produce around $\mathcal{O}(10)$ $\phi$ particles in the final state (through $\phi \to \phi \phi$ splitting). Both $\phi$ and $\chi$ for benchmark point ($iv$) will produce a few $\phi$ particles in the final state (through $\chi \to \phi \chi$ splitting). 
This leads to lepton-jet (LJ) signature~\cite{Arkani-Hamed:2008kxc,Baumgart:2009tn,Chan:2011aa,Buschmann:2015awa} for each of the initial $\phi$ and $\chi$. The ATLAS collaboration has searched for both prompt and displaced lepton jets~\cite{ATLAS:2015itk,ATLAS:2019tkk}, aiming to the mediator mass $\gtrsim 0.1$ GeV. Moreover, the mediator multiplicity in each lepton jet is not larger than four. So those existing searches are not optimal for our benchmark points. 
For benchmark point ($ii$) and ($iii$), the $\phi \to \phi \phi$ splitting is so copious such that the transverse momenta of final state leptons are below the selection threshold in lepton jet searches (which requires energy of electron to be greater than 10 GeV). 
As a result, the signatures of $\phi$ and $\chi$ become Soft Unclustered Energy Patterns (SUEP). There have been some studies on searching for SUEP~\cite{Harnik:2008ax,Knapen:2016hky,Barron:2021btf}, focusing on hadronic final states. 
The detailed study for the collider search of our benchmark points will be addressed in future works. 

\begin{table}[t!]
\centering
\begin{tabular}{|c|cc|cc|cc|cc|}  
\hline
Benchmark point & \multicolumn{2}{|c|}{(i)} & \multicolumn{2}{|c|}{(ii)} & \multicolumn{2}{|c|}{(iii)} & \multicolumn{2}{|c|}{(iv)} \\ \hline
Particle from Higgsino decay & $\phi$ & $\chi$ & $\phi$ & $\chi$ & $\phi$ & $\chi$ & $\phi$ & $\chi$  \\\hline
Signature & LJ & $\slashed{E}_T$ & SUEP & $\slashed{E}_T$  & SUEP & SUEP & LJ & LJ  \\ \hline
\end{tabular}
\caption{\label{tab:signa} The signatures of $\chi$ and $\phi$ (from the Higgsino decay) for benchmark points}
\end{table}

\section{Conclusion}

We study the SIDM scenario in the general NMSSM and beyond, where the DM is a Majorana fermion and the force mediator is a scalar boson. Due to the relatively large couplings and light scalar mediator in this scenario, the DM/mediator will go through multiple branchings if they are produced with high energy, leading to the signature of multiple scalars in the final state. 

The DM self-interaction cross section is calculated in both analytical and numerical ways. 
In particular, an improved analytical estimation for the DM self-interacting cross section which takes into account the Born level effects is proposed. 
Based on the scanned points in the NMSSM, we demonstrate that the analytical expression matches well with the numerical solution. In most case, the relative difference of DM self-interacting cross section between two methods is $\sim \mathcal{O} (10)\%$.

Two benchmark points in the general NMSSM that address the small scale structure problem with correct DM relic density and satisfy other phenomenological constraints are explicitly given for the first time (one of the benchmark points is challenged by the DM direct searches). 
They are featured by relatively small $\mu$ parameter, light singlet-like scalar mediator and relatively large triple scalar interaction. 
In order to have enough DM relic density, $\kappa \gtrsim 0.2$ is not allowed in the NMSSM if the DM is lighter than 20 GeV. Otherwise, the annihilation of $\chi \chi \to \phi \phi$ will dilute DM density efficiently. 
The branching ratio of the lighter neutral Higgsino decaying into the singlet scalar and singlino is proportional to $\kappa^2$, which is around $10^{-3}$. 
The production rates of the Higgsinos pair at the 13 TeV LHC with at least one of the Higgsino decaying into the singlet scalar and singlino are 0.108 fb and 8.95 fb for benchmark point ($i$) and ($ii$), respectively.

A Monte Carlo simulation of the DM/mediator showers are implemented by building the multiple branchings as a Markov process based on the Sudakov form factors. 
For the two benchmark points in the NMSSM, $\kappa \sim 0.1$, which means the splitting of $\chi \to \phi \chi$ is suppressed. On the other hand, the splitting probability of $\phi \to \phi \phi$ is high. 
The number of $\phi$ (with highest probability) from the Higgsino production and decay at the LHC can reach $\sim 20$ for benchmark point ($i$) and is even more for benchmark point ($ii$). 
Two benchmark points beyond the NMSSM with $\kappa=2.5$ and address the small scale structure issue are proposed as well, to illustrate the cases with sizable $\chi \to \phi \chi$ splitting. 
When $\phi \to \phi \phi$ is turned off (corresponding to benchmark point ($iv$)), number of $\phi$ (with highest probability) in the final state is around $3\sim 4$. 
The $p_T$ spectra of DM/mediator are also shown for benchmark points. 

Finally, we comment on the tension between the Big Bang Nucleosynthesis (BBN) limits and dark matter direct detection constraints for the SIDM scenario with light long-lived mediator. Extensions to the simple singlet-dominant DM sector are required to alleviate the tension. 
Depending on the form of possible extension, the energetic $\phi$ and $\chi$ from Higgsino decay can induce remarkable signatures at the LHC.


\begin{acknowledgments}
This work was supported in part by the Fundamental Research Funds for the Central Universities, by the National Science Foundation of China under Grant No. 11905149, by Projects No. 11847612 and No. 11875062 supported by the National Natural Science Foundation of China, and by the Key Research Program of Frontier Science, Chinese Academy of Sciences.

\end{acknowledgments}

\bibliographystyle{jhep}
\bibliography{nmssm_simp}
\end{document}